\documentclass[aps,pre,showpacs,floatfix]{revtex4}
\usepackage{graphicx}
\usepackage{amsmath}
\usepackage{amssymb}
\usepackage{color}
\usepackage[utf8]{inputenc}
\usepackage{verbatim}
\usepackage{overpic}

\usepackage[tight,sf]{subfigure}%

\newcommand{\romd}{{\text{d}}}

\newcommand{\romT}{{\text{T}}}

\newcommand{\VECe}{{\boldsymbol{e}}}
\newcommand{\VECf}{{\boldsymbol{f}}}
\newcommand{\VECl}{{\boldsymbol{l}}}
\newcommand{\VECn}{{\boldsymbol{n}}}

\newcommand{\VECx}{{\boldsymbol{x}}}
\newcommand{\VECy}{{\boldsymbol{y}}}
\newcommand{\VECz}{{\boldsymbol{z}}}

\newcommand{\VECX}{{\boldsymbol{X}}}

\newcommand{\CALE}{{\mathcal{E}}}
\newcommand{\CALK}{{\mathcal{K}}}

\newcommand{\VECrho}{{\boldsymbol{\rho}}}
\newcommand{\VECLambda}{{\boldsymbol{\Lambda}}}

\newcommand{\RR}{\mathbb{R}}

\begin{document}
\title{Toroidal membrane vesicles in spherical confinement}

\author{ Lila Bouzar$^{1}$, Ferhat Menas$^{2,3}$, and Martin Michael M\"uller$^{4,5}$} 

\affiliation{
$^1$ D\'epartement de Physique Th\'eorique, Facult\'e de Physique, USTHB; BP~32 El-Alia Bab-Ezzouar, 16111 Alger, Algeria\\
$^2$ Laboratoire de Physique et Chimie Quantique, Universit\'e Mouloud Mammeri; BP~17, 15000 Tizi-Ouzou, Algeria\\
$^3$ Ecole Nationale Pr\'eparatoire aux Etudes d'Ing\'eniorat, D\'ep.~de Phys.; BP~05 Rou{\"i}ba, Alger,  Algeria\\
$^4$ Equipe BioPhysStat, LCP-A2MC, Universit\'{e} de Lorraine; 1 boulevard Arago, 57070 Metz,  France\\
$^5$ Institut Charles Sadron, CNRS-UdS; 23 rue du Loess, BP 84047, 67034 Strasbourg cedex 2, France
} 
\date{\today}
   
\begin{abstract}
We investigate the morphology of a toroidal fluid membrane vesicle confined inside a spherical container. 
The equilibrium shapes are assembled in a geometrical phase diagram as a function of scaled area and 
reduced volume of the membrane. For small area the vesicle can adopt its free form. When increasing 
the area, the membrane cannot avoid contact and touches the confining sphere along a circular contact line, 
which extends to a zone of contact for higher area. The elastic energies of the equilibrium shapes are 
compared to those of their confined counterparts of spherical topology to predict under which conditions a 
topology change is favored energetically. 
\end{abstract}
 
% 87.16.D-: Membranes, bilayers, and vesicles
% 87.10.Pq: Elasticity theory

\maketitle

%%%%%%%%%%%%%%%%%%%%%%%%%%%%%%%%%%%%%%%%%%%%%%%%%%%%%%%%%%%%%%%%%%%%%%%%%%%%%%%%

\section{Introduction \label{sec:introduction}}
Topological transitions abound in physical and biological systems. In fluid dynamics they can be induced by capillary instabilities like   
the Plateau---Rayleigh instability which causes a fluid thread to break up into smaller droplets \cite{Eggers1997}.  Another classic example 
concerns area-minimizing surfaces: a catenoidal soap film spanning two circular wires undergoes a transition to two soap films covering both wires separately when the distance between the wires is increased \cite{Isenberg1992}. A topological transition can also be induced by deforming the wire such as in the case of a M{\"o}bius strip which can change its topology with a twist singularity \cite{goldstein2010}.

In the biological world, cells are controlling the topology of their ingredients at all length scales. At the molecular level the function of DNA \cite{Wang1996} and proteins \cite{Clementi2000} depends crucially on their topology. On a larger scale membrane-bound organelles such as the endoplasmic reticulum or mitochondria display intricate geometries whose topological features are carefully tuned for proper function \cite{Voeltz2002,mannella2006,Terasaki2013,Guven2014}. A healthy cell can change the shape and topology of its membranes with the help of other constituents such as the cytoskeleton \cite{Fletcher2010,Brown2005}. However, topology changes can also occur due to pathological conditions: 
one example is X-linked centronuclear myopathy---a congenital muscle weakness---in which the membrane stacks of the sarcoplasmic reticulum of skeletal muscle cells are remodeled into cubic membrane structures of higher genus \cite{Amoasii2013}. Similar structures have been observed in the inner membrane of mitochondria upon starvation \cite{zakaria2009}. 

A theoretical ansatz to model the morphology of the mitochondrium consists of studying a fluid membrane vesicle inside a confining spherical container \cite{OsmanNorbertMartin2012A,OsmanNorbertMartin2012B,Sakashita2014,DePascalis2014,Rim2014,Guven2013A}. Although this model is too simple to capture all the details of the complex folding patterns, one obtains the basic morphological building blocks. Experimental observations \textit{in vivo} \cite{john2005} and in artificial systems \cite{Sakashita2014} confirm the theoretical predictions. 
However, the spherical topology of the confined membrane vesicle has been fixed in all existing theoretical studies. 

To handle topology changes as well, we will consider confined fluid membrane vesicles of toroidal topology. The equilibrium shapes of free toroidal vesicles without confinement have been studied since the 1990s \cite{OuYang1990,Seifert1990,Seifert1991A,MutzBensimon1991,Fourcade1992A,Fourcade1992B,Michalet1995,Juelicher1993B}.
Only recently new stable non-axisymmetric shapes have been found \cite{Sakashita2015}. 
As in these articles we use the classical curvature model to study our system (see Sec.~\ref{sec:model}). To obtain a geometrical phase diagram with all equilibrium shapes, we start by analyzing the free solutions and determine the parameter values at which the vesicle starts to touch the confining sphere (Sec.~\ref{sec:freesolutions}). In the next step we determine the axisymmetric  (Sec.~\ref{sec:confinedaxisymmetricsolutions}) and non-axisymmetric (Sec.~\ref{sec:confinednonaxisymmetricsolutions}) equilibrium shapes in contact with the confinement. In Sec.~\ref{sec:comparisonsphericaltopology} we finally compare the vesicles of spherical and toroidal topology to predict under which conditions a topological transition can take place.

%%%%%%%%%%%%%%%%%%%%%%%%%%%%%%%%%%%%%%%%%%%%%%%%%%%%%%%%%%%%%%%%%%%%%%%%%%%%%%%%

\section{Model \label{sec:model}}

%%%%%
\begin{figure}[t]
\centering
\includegraphics*[width=0.65\textwidth,bb= 0 70 480 495]{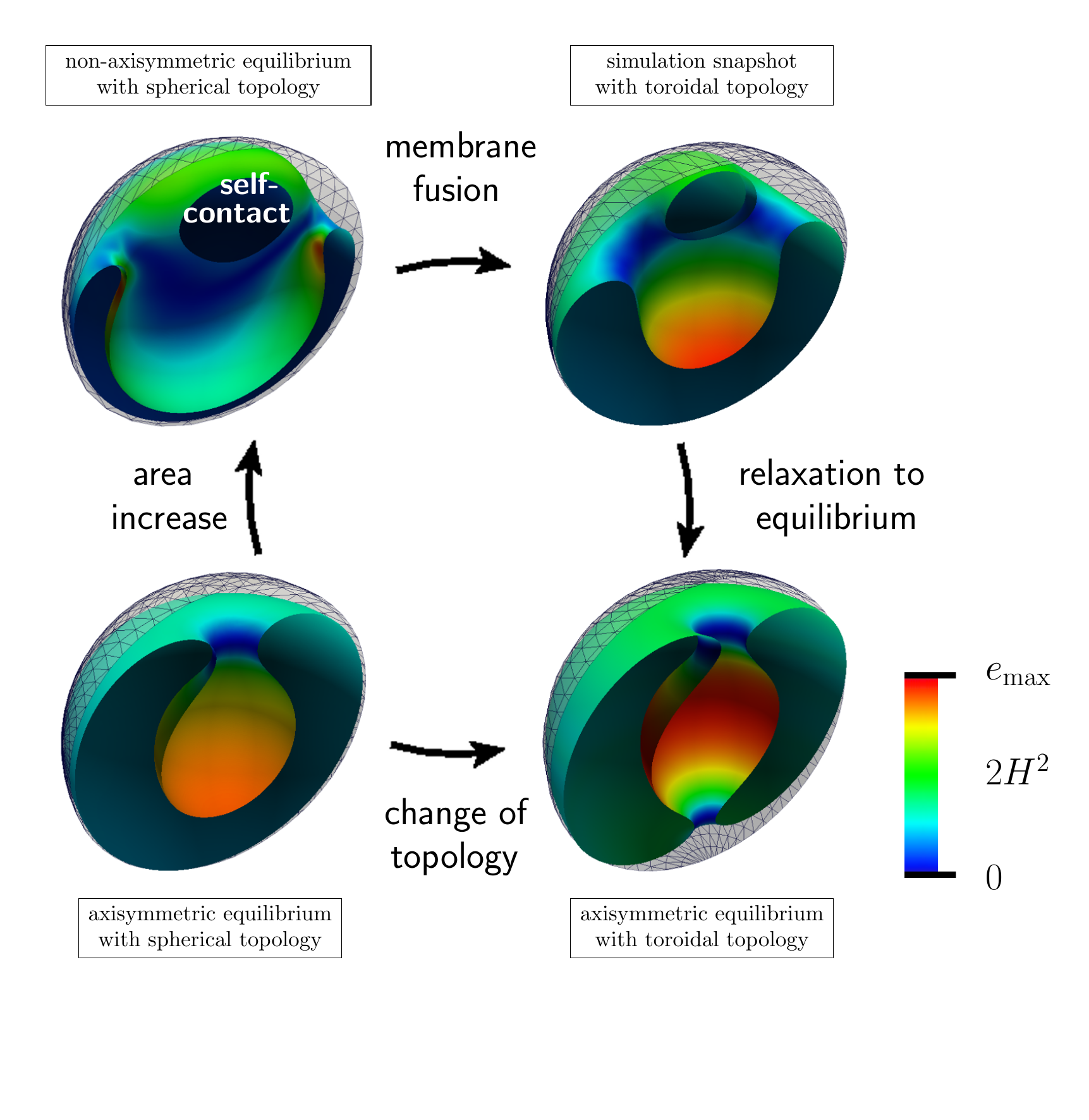}
\caption{(Color online) Transition from a spherical to a toroidal topology (see text): 
(\textit{bottom left}) axisymmetric equilibrium solution with spherical topology $(a,v)=(1.2,0.8)$, 
(\textit{top left}) ellipsoid-like equilibrium solution with spherical topology $(a,v)=(1.5,0.8)$, 
(\textit{top right}) confined toroidal membrane $(a,v)=(1.2,0.8)$ (simulation snapshot), and 
(\textit{bottom right}) axisymmetric equilibrium solution with toroidal topology $(a,v)=(1.2,0.8)$. 
The maximum values of the scale bar are $e_\text{max}= 10$, $23$, $11$, and $8$ (clockwise starting from bottom left). 
}%
\label{fig:topologychange}%
\end{figure}
%%%%%
We search for the equilibrium shapes of a toroidal membrane vesicle of fixed area $\bar{A}$ and volume $\bar{V}$ in spherical confinement. In the following area and volume are scaled by the respective quantities $A_0$ and $V_0$ of the confining sphere: 
$a=\bar{A}/A_0$ and $v=\bar{V}/V_0$.
The elasticity of the membrane is modeled within the scope of the classical curvature model \cite{Canham,Helfrich,Evans} with the bending energy
\begin{equation}
  E_b := E_\kappa + E_{\bar{\kappa}} = \frac{\kappa}{2} \int \romd A \; (2 H - C_0)^2 + \bar{\kappa} \int \romd A \; K \;  ,
  \label{eq:bendingenergy}
\end{equation}
where $H$ is the mean curvature and $K$ the Gaussian curvature. The stiffness of the membrane is reflected in the material parameters $\kappa$ and $\bar{\kappa}$ which are the bending and the Gaussian rigidity, respectively. We assume in particular that the membrane prefers to be intrinsically flat which implies that the spontaneous curvature $C_0$ is zero in our study. 
The second term of the energy involving the Gaussian curvature is a topological invariant for a closed surface. According to the Gauss-Bonnet theorem it is given by \cite{Kreyszig}
\begin{equation}
   E_{\bar{\kappa}} = \bar{\kappa} \int \romd A \; K = 4\pi \bar{\kappa} (1 - g)
\end{equation}
for a surface of genus $g$, where $g$ corresponds to the number of ``handles'' of the vesicle. For a spherical surface such as those studied in \cite{OsmanNorbertMartin2012A, OsmanNorbertMartin2012B} $g$ is zero; for the toroidal surfaces considered here $g$ is equal to one and $E_{\bar{\kappa}}$ vanishes. Depending on the sign of $\bar{\kappa}$ an increase of $g$ will thus correspond to an abrupt increase or decrease in energy, respectively. For fluid lipid mono- and bilayers one typically finds $\bar{\kappa}\approx -\kappa < 0$ \cite{Hu2012}, which implies that a change from a spherical to a toroidal topology will increase the term of the energy involving $K$. However, the first term of the energy will change as well and it is not clear \textit{a priori} whether the total energy will increase or decrease. 

Why should the topology of the membrane change at all?  The equilibrium shapes of a \emph{spherical} membrane vesicle in confinement have been determined recently by some of the authors \cite{OsmanNorbertMartin2012A, OsmanNorbertMartin2012B}. For moderate values of area ($a>1$) we have found an axisymmetric solution with a ``light bulb''-like invagination which starts to contact itself when the area of the vesicle is increased (see left part of Fig.~\ref{fig:topologychange}). If the membrane was able to fuse, for instance, with the help of embedded specialized molecules such as SNARE proteins \cite{Jahn2006}, the two parts of the surface in contact could merge. The topology of the resulting surface is toroidal (see right part of Fig.~\ref{fig:topologychange}): two necks instead of one connect the invagation to the outer part of the membrane in contact with the spherical container. The toroidal ground state for the parameters of Fig.~\ref{fig:topologychange} is axisymmetric. We will see in the following that the symmetry of the solution depends on the respective values of area and volume of the vesicle.

%%%%%%%%%%%%%%%%%%%%%%%%%%%%%%%%%%%%%%%%%%%%%%%%%%%%%%%%%%%%%%%%%%%%%%%%%%%%%%%%

\section{Free solutions \label{sec:freesolutions}}

%%%%%
\setlength{\fboxsep}{0.5mm} % distance number <> box in figure caption
\begin{figure}[t]
\centering
\includegraphics[width=0.45\textwidth]{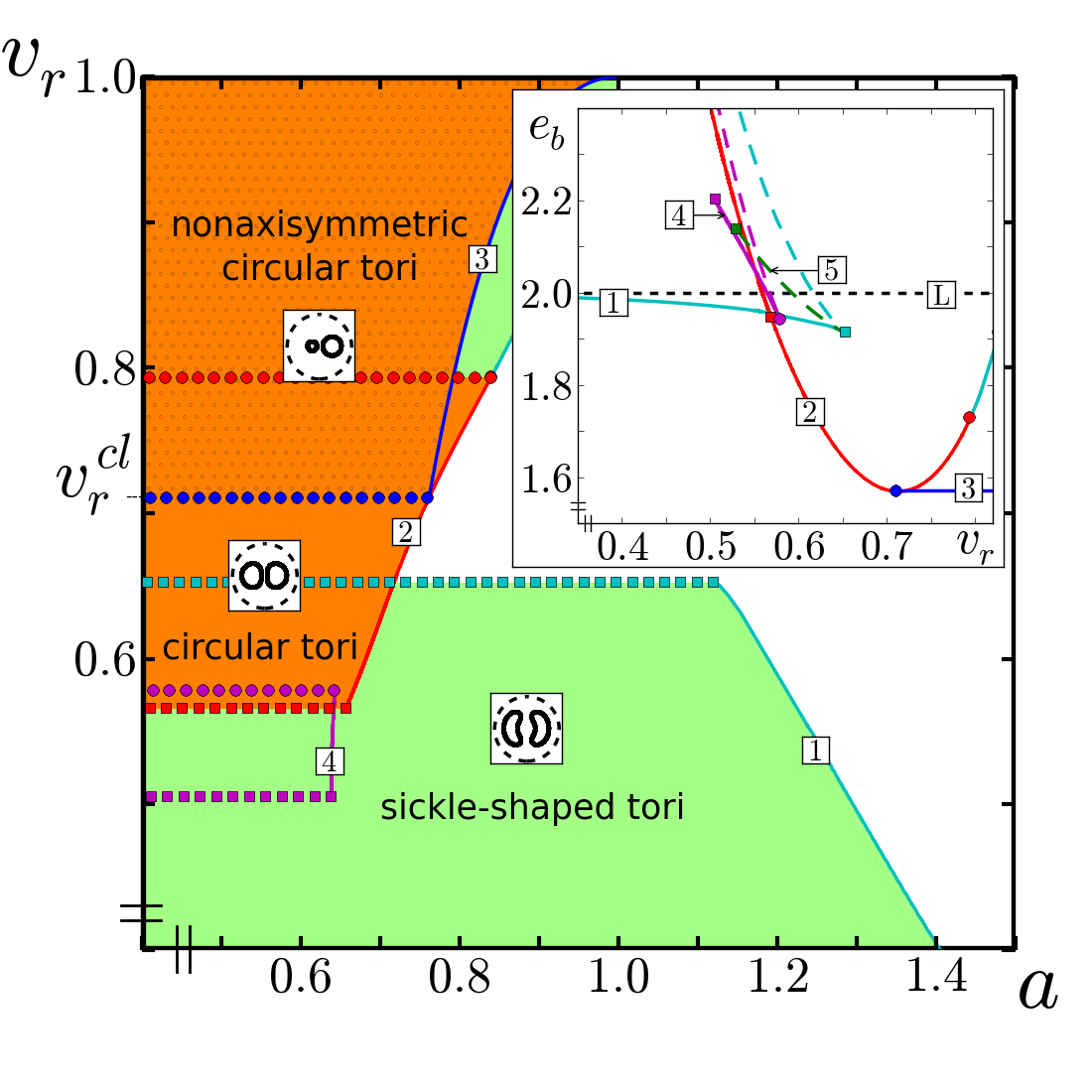}
\caption{(Color online)
Partial phase diagram $(a,v_r)$ showing the free axisymmetric equilibrium shapes together with the conformal transformations of the Clifford torus. The colored regions indicate the \emph{global} minima of this set of shapes, which are either sickle-shaped tori (lime green region), circular tori (orange region), or conformal transformations of the Clifford torus (hatched orange region). The discoid tori (purple) are local minima of higher energy. The curves \fbox{1} - \fbox{4} delimit the regions in which the free shapes are equilibrium solutions of the confinement problem. To their right no such solution exists, which explains why parts of the phase diagram are white.  
\textit{Inset:} Scaled bending energy $e_b:=E_b/(8\pi\kappa)$ as a function of $v_r$ for the different solution types (see Ref.~\cite{Seifert1991A,Juelicher1993B}): sickle-shaped (lime green curve \fbox{1}), circular (red curve \fbox{2}), discoid (purple curve \fbox{4}), and stomatocyte (dark green curve \fbox{5}) tori, and the conformal transformations of the Clifford torus (blue curve \fbox{3}). The long-dashed parts of these curves correspond to shapes of maximum bending energy. The short-dashed line \fbox{L} corresponds to the limit torus with $e_b=2$, which consists of two identical spheres connected \textit{via} two infinitesimal holes.}
\label{fig:phasediagramfreesolutions}%
\end{figure}
%%%%%

Because of the scale invariance of the bending energy the only relevant parameters of the free vesicle are the reduced volume $v_r=\frac{v}{a^{3/2}}$ and the spontaneous curvature $C_0$. For vanishing spontaneous curvature the following types of axisymmetric local extrema have been found (see Refs.~\cite{Seifert1990, Seifert1991A,Juelicher1993B} and Fig.~\ref{fig:phasediagramfreesolutions}): $(i)$ sickle-shaped tori, $(ii)$ (almost) circular tori, 
$(iii)$ discoid tori, and $(iv)$ toroidal stomatocytes. 
All but the stomatocytes have an additional reflection symmetry with respect to the plane perpendicular to the axis of symmetry. 

The lower limits of discoid tori and toroidal stomatocytes are shapes with a vanishing hole diameter at the vertical axis (purple and dark green squares in Fig.~\ref{fig:phasediagramfreesolutions}). The lower limit of the stable sickle-shaped tori, $v_r\to 0$, corresponds to two connected identical spheres with an infinitesimal hole in each of the poles. In the same limit the (almost) circular tori approach a shape of exact circular 
cross section while both the diameter of the hole and the cross section vanish \cite{Seifert1991A}.  
At $v_r=v_r^{cl} = \frac{3}{2^{5/4}\pi^{1/2}}\approx 0.71$ the cross section of the corresponding profile of the same branch is exactly circular as  well (blue circles in Fig.~\ref{fig:phasediagramfreesolutions}). This is the Clifford torus whose bending energy is a lower bound to that of any toroidal shape \cite{Marques14}. Its two different ellipsoidal deformations have been identified as discoid $v_r<v_r^{cl}$ and sickle-shaped $v_r>v_r^{cl}$ tori \cite{Juelicher1993B}. In this article we will use a slightly different definition by calling only those tori sickle-shaped, which \textit{have} a sickle-shaped cross section, $i.e.$, a cross-sectional meridional curvature which changes sign twice. With this definition the branch of circular tori becomes sickle-shaped for $v_r\ge0.79$ (red circles in Fig.~\ref{fig:phasediagramfreesolutions}). 
All other solutions of this branch (and the corresponding solutions in partial or full contact with the container, see below) will be called circular tori in the following. 

In the inset of Fig.~\ref{fig:phasediagramfreesolutions} the scaled bending energy $e_b:=E_b/(8\pi\kappa)$ is shown as a function of $v_r$ \cite{Seifert1990}. The branch \fbox{5} of the toroidal stomatocytes contains no minima and will thus be omitted in the following. Without the confinement constraint the global minima can be directly read off: for $v_r<0.57$ the tori are sickle-shaped. Above this value the circular tori have a lower bending energy. Increasing $v_r$ leads to a discontinuous shape transformation (red squares in Fig.~\ref{fig:phasediagramfreesolutions}) since the branch of locally stable sickle-shaped tori runs up to $v_r=0.65$ (lime green squares). 
For $v_r>v_r^{cl}$ the ground states are non-axisymmetric shapes. They can be constructed with the help of conformal transformations \cite{Seifert1990}.

Special conformal transformations of the form
\begin{equation}
  \VECX' = \frac{\frac{\VECX}{\VECX^2} + \VECLambda}{(\frac{\VECX}{\VECX^2} + \VECLambda)^2}
  \; .
  \label{eq:specialconformaltransformation}
\end{equation}
conserve the toroidal topology of the surface $\VECX (\xi^1,\xi^2)\in\RR^3$ and leave the bending energy invariant. 
They consist of an inversion about the origin, followed by a translation along a constant vector $\VECLambda$, and a second inversion about the origin. The resulting surface $\VECX' (\xi^1,\xi^2)$ has a reduced volume which is larger or equal to the reduced volume of the original surface.  $\VECX'$ can always be rescaled to stay inside the container. 
Above $v_r^{cl}$ the global minima are thus non-axisymmetric conformal transformations of the Clifford torus since their bending energy $e_b=\pi / 2$ (blue line \fbox{3} in the inset of Fig.~\ref{fig:phasediagramfreesolutions}) is always lower than that of the axisymmetric solution of same reduced volume (red curve \fbox{2} in the inset of Fig.~\ref{fig:phasediagramfreesolutions}) \cite{Seifert1990, Seifert1991A,Juelicher1993B}. 
The most general special conformal transformation~(\ref{eq:specialconformaltransformation}) involves the three parameters $\VECLambda=(\Lambda_x,\Lambda_y,\Lambda_z)$. For an axisymmetric surface $\VECX$ only two parameters remain: without loss of generality we can orient $\VECLambda$ to lie in the plane defined by the axis of symmetry $\VECz$ and the perpendicular axis $\VECx$. Since conformal transformations of the Clifford torus along the axis of symmetry rescale the shape, there is only a one-parameter family of special conformal transformations leading to the global non-axisymmetric minima discussed above \cite{Juelicher1993B}. 

As long as the area of the vesicle is small enough, the confined vesicle does not touch the container at all and adopts the shape of the free system. To predict the behavior for larger area, one can rescale the free axisymmetric solutions up to the point at which they come into contact with the spherical container. These solutions have been obtained numerically with a shooting method (see appendix~\ref{app:RungeKutta} for the details of the method). They correspond to the curves \fbox{1} - \fbox{4} delimiting the parts in the phase diagram $(a,v_r)$  in which the free vesicles are the local minima (see Fig.~\ref{fig:phasediagramfreesolutions}). To understand how the \emph{global} energy minima (colored regions of Fig.~\ref{fig:phasediagramfreesolutions}) are found, consider, for example $v_r = 0.6$. Without the confinement the free circular torus is the global minimum whereas the free sickle-shaped torus is a local minimum. At $a = 0.6$ both solutions can be put inside the container without any contact between membrane and confinement. However, when we increase the parameter $a$ keeping $v_r$ constant such that we cross the red line \fbox{2} in Fig.~\ref{fig:phasediagramfreesolutions}, we do not find any circular tori which are \emph{not} in contact with the confinement. The only free solution is the corresponding sickle-shaped torus which is thus the global and only minimum of all free shapes. In general, one observes that the sickle-shaped tori start contacting the confinement for an area which is much larger than the corresponding value for the circular tori. Consequently, the sickle-shaped tori are the only free minima for $0<v_r<0.65$ provided the area is large enough. In contrast to that we can neglect the locally stable discoid solutions for the discussion in the following since they start contacting the container for even smaller area than the circular tori. 

To find the delimiting curve of the region of the free non-axisymmetric solutions, $i.e.$, the blue curve \fbox{3} of the phase diagram in Fig.~\ref{fig:phasediagramfreesolutions}, 
we consider the Clifford torus in contact in the horizontal ($\VECx\VECy$) plane. Its defining radii are given by $r=\frac{1}{\sqrt{2}+1}$ and $R=\sqrt{2}r$. In polar coordinates $(\theta,\phi)$ its shape can be written as:
\begin{equation}
  \VECX(\theta,\phi) = \frac{1}{\sqrt{2}+1} ( (\sqrt{2}+\sin{\theta})\cos{\phi}, (\sqrt{2}+\sin{\theta})\sin{\phi}, \cos{\theta} )
    \; ,
\end{equation}
on which we perform the conformal transformation, Eq.~(\ref{eq:specialconformaltransformation}). With an appropriate rescaling and a final translation along $\VECLambda$
\begin{equation}
  \bar{\VECX} = \VECX' (1-\Lambda^2) + \VECLambda
    \; ,
  \label{eq:conformaltransformationconfined}
\end{equation}
every point of the inner membrane which lies on the confining sphere will stay on the sphere. Choosing $\VECLambda=\Lambda_x \VECx$ 
one finds a surface with the same circular contact line as the original Clifford torus.  In the limit $\Lambda_x\to 1$, ($a,v_r)\to (1,1)$ and the conformally transformed torus becomes a sphere with an infinitesimal handle. The resulting shapes are Dupin cyclides whose centers of curvature are located on two confocal conics: an ellipse and a hyperbola \cite{Fourcade1992B,Kleman1977}. For the Clifford torus the ellipse corresponds to a circle of radius $R$ and the hyperbola becomes a line which coincides with the axis of symmetry. Using the results of Refs.~\cite{Fourcade1992B,Kleman1977} one obtains for the area and volume of the transformed torus:
\begin{subequations}
\begin{eqnarray}
  a & = & 4 (1-\lambda) \lambda [\CALE (e^2) + \frac{e^2-1}{2} \CALK (e^2)]
  \; , \quad \text{and}
  \\
  v & = & 8 \lambda^3 [\frac{(1-\lambda)^2}{\lambda^2} \CALE (e^2) +  \frac{e^2-1}{2} (1+\frac{3}{4}\frac{e^2-1}{2})\, \CALK (e^2)]
  \; ,
\end{eqnarray}\label{eq:touchingtransformedCliffordtorus}%
\end{subequations}
where $\CALK (e^2)$ and $\CALE (e^2)$ are the complete elliptic integrals of the first and second kind, respectively \cite{Abramowitz}. The parameter 
$e = \sqrt{\frac{ \lambda^2- 4 \lambda + 2}{ \lambda^2}}\in \{0,1\}$ is the eccentricity and $\lambda=\frac{2+\sqrt{2} + (-2 + \sqrt{2}) \Lambda_x^2}{3+2 \sqrt{2} + (-3 + 2\sqrt{2}) \Lambda_x^2}\in \{0.5,R\}$ is the semi-major axis of the generating ellipse. 

Fig.~\ref{fig:phasediagramfreesolutions} does not show the whole story yet. 
We have not yet discussed the conformal transformations of the sickle-shaped tori. % and the tori of the circular branch for  $v_r>v_r^{cl}$. 
The resulting surfaces have to be compared energetically to the solutions which are in contact with the confinement. We will do this in two steps. We first identify all axisymmetric solutions of the confinement problem. In a second step we determine the non-axisymmetric confined shapes and assemble everything in one phase diagram.

%%%%%%%%%%%%%%%%%%%%%%%%%%%%%%%%%%%%%%%%%%%%%%%%%%%%%%%%%%%%%%%%%%%%%%%%%%%%%%%%

\section{Confined axisymmetric solutions \label{sec:confinedaxisymmetricsolutions}}

%%%%%
\begin{figure}[t]
\centering
\includegraphics[width=0.99\textwidth]{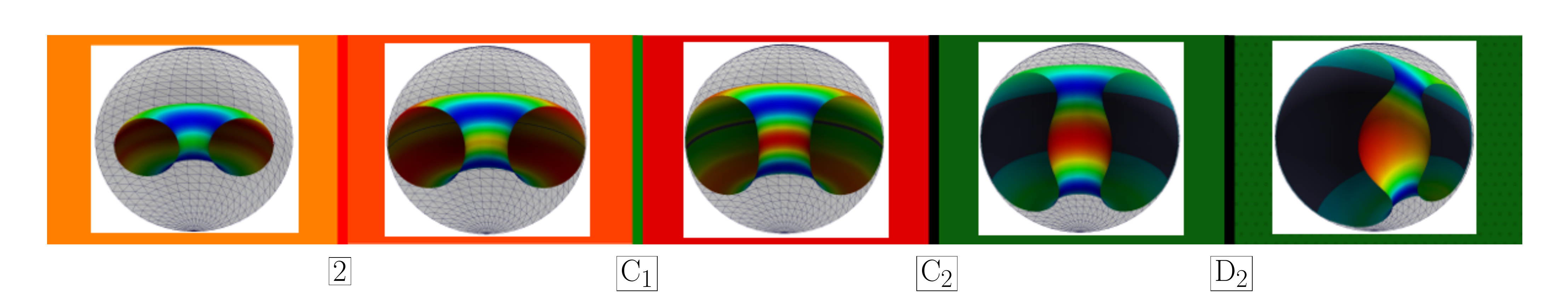}
\caption{(Color online) Equilibrium profiles for fixed reduced volume $v_r=0.7$ and increasing area (from left to right) $a=0.5$, $0.8$, $0.9$, $1.1$, and $1.15$. The background colors and labels are in accordance with the final phase diagram (Fig.~\ref{fig:phasediagramfinal}). The color code of the profiles is the same as in Fig.~\ref{fig:topologychange} with $e_\text{max}=10$, $5.5$, $6$, $10$, and $10.5$.}
\label{fig:profiles}%
\end{figure}
%%%%%

When you press a cylindrical sheet of paper gently against a wall, the contact between the two will be a line. The curvature of the sheet perpendicular to the line of contact will generally not be zero. Increasing the force will decrease this curvature until it is equal to zero, $i.e.$, equal to the curvature of the flat wall. Above this threshold a contact area will start to form. 
The same phenomenon is observed in our system (see the first four profiles in Fig.~\ref{fig:profiles} for an example): increasing the area for a fixed reduced volume one reaches a point at which the free vesicle starts to touch the spherical confinement. The corresponding curves in the phase diagram $(a,v_r)$ have been discussed in the previous section. Interestingly, all these solutions are in contact with the container in one circle which lies in the horizontal symmetry plane. Their curvature perpendicular to the contact line, $K_\perp$, is always larger than that of the sphere, $\tilde{K}_\perp=1$ . Increasing the area further thus leads to solutions which are still confined by the circle (called \emph{contact-circle solutions} in the following). Above a certain threshold area, $K_\perp=\tilde{K}_\perp$, and the circular contact line becomes a spherical segment leading to \emph{contact-area solutions}. 

%%%%%
\begin{figure}[t]
\centering
\subfigure[][]{\label{fig:phasediagramaxi_a}\includegraphics[width=0.48\textwidth]{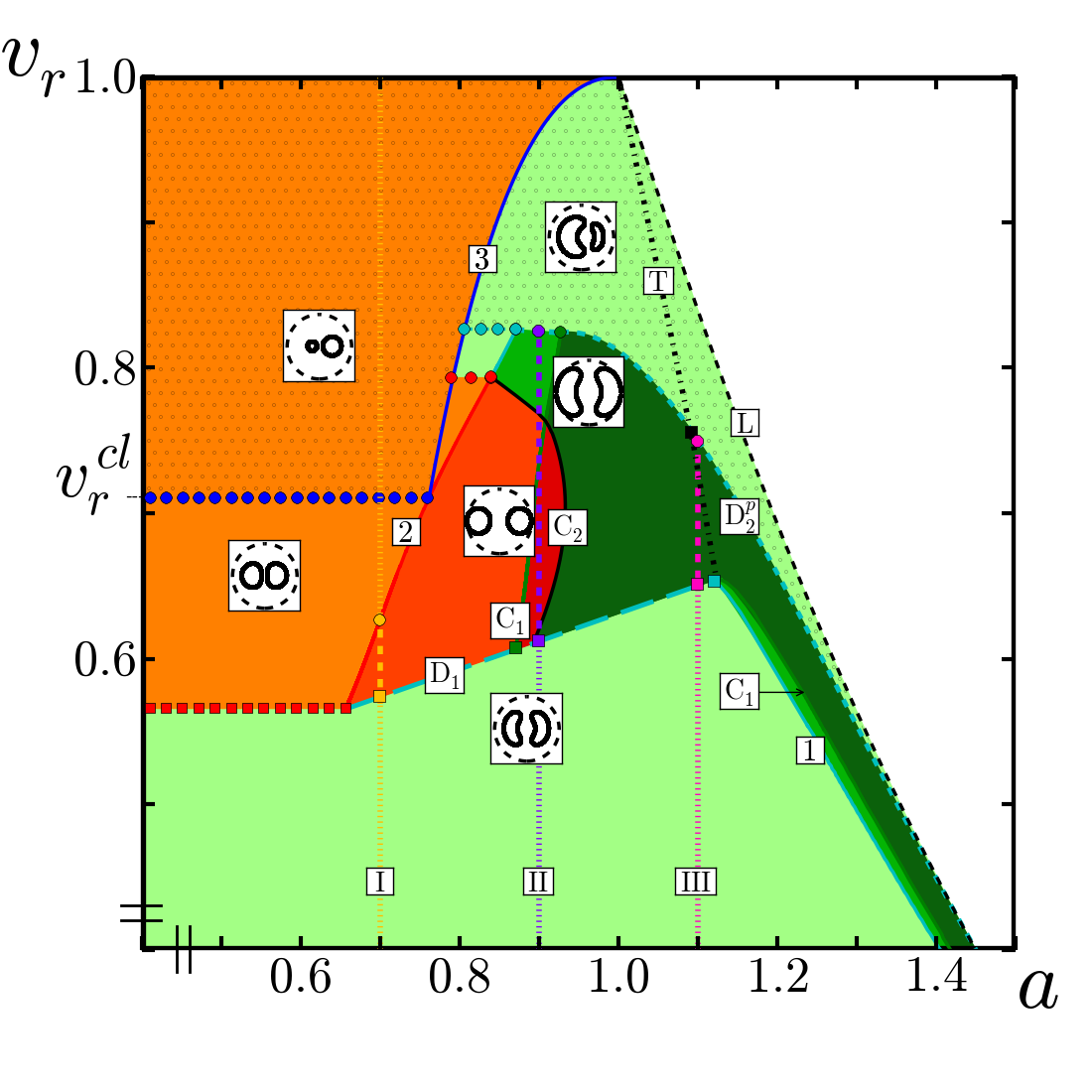}}
\hfill
\subfigure[][]{\label{fig:phasediagramaxi_b}\includegraphics[width=0.48\textwidth]{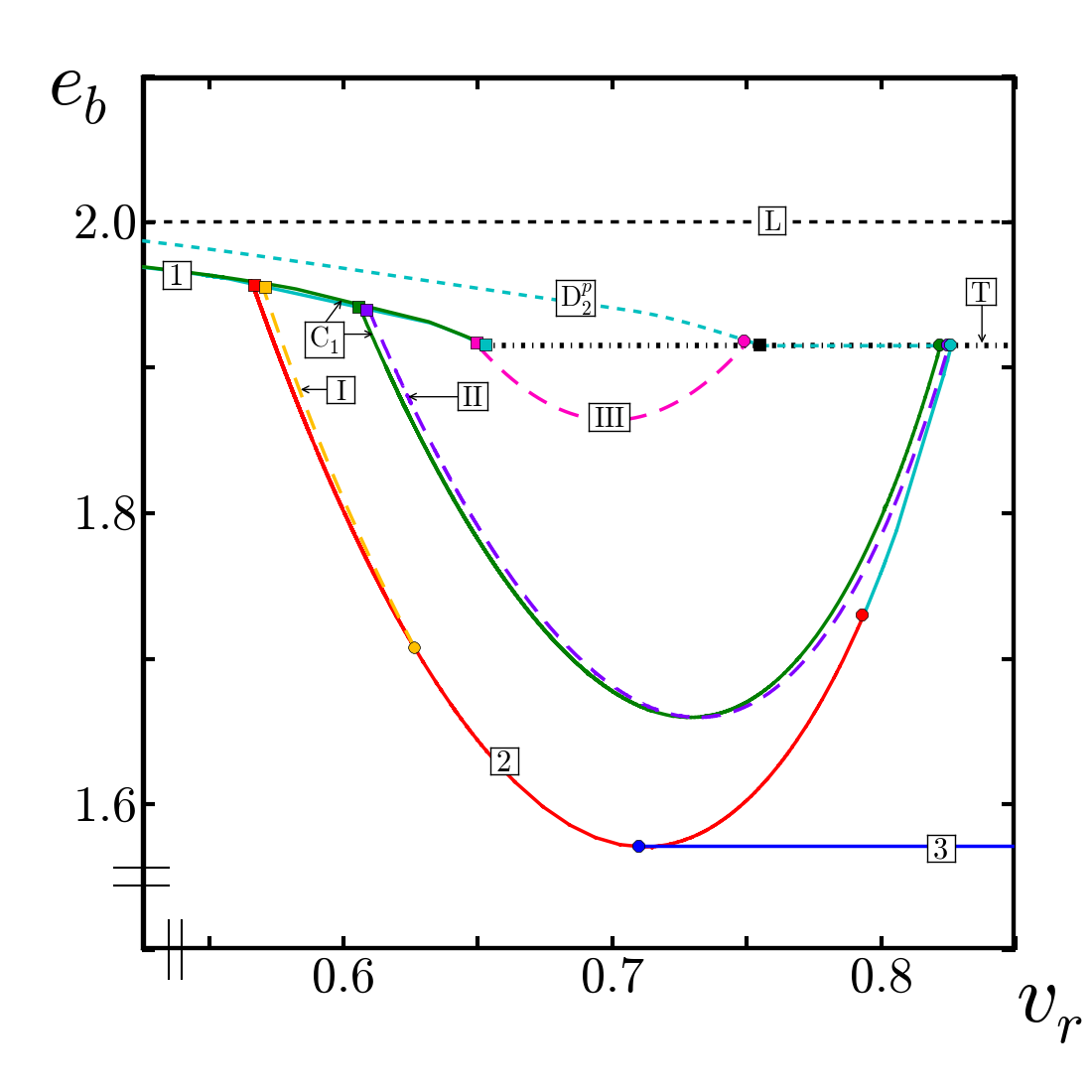}}
\caption{(Color online) (a) Partial phase diagram $(a,v_r)$ showing the axisymmetric extrema together with the conformal transformations of the free vesicles. The colored regions indicate the \emph{global} minima of this set of shapes. Regions of sickle-shaped tori are shown in green, regions of circular tori in orange/red.  Non-axisymmetric regions are hatched. The colors vary from light to dark with increasing confinement. The curve \fbox{$\mathrm{C}_1$} separates the region of contact-circle solutions from the region of contact-area solutions, whereas \fbox{$\mathrm{C}_2$} is the boundary between the confined circular and sickle-shaped solutions (for a description of the curves \fbox{$\mathrm{D}_1$}, \fbox{$\mathrm{D}_2^p$}, and \fbox{T} see text).
(b)  Scaled bending energy $e_b$ as a function of $v_r$ for the different axisymmetric solutions. In addition to the free solutions (solid lime green \fbox{1}, red \fbox{2}, and blue curves \fbox{3}) lines of constant area $a$ are shown as well (dashed yellow \fbox{I} ($a=0.7$), purple   \fbox{II} ($a=0.9$), and pink  \fbox{III} ($a=1.1$) curves). The short-dashed line \fbox{L} corresponds to the limit torus with $e_b=2$, which consists of two identical spheres connected \textit{via} two infinitesimal holes.}%
\label{fig:phasediagramaxi}%
\end{figure}
%%%%%
The corresponding confined axisymmetric solutions have been determined numerically with the same shooting method as in the previous section (see appendix~\ref{app:RungeKutta}). The appropriate boundary conditions depend on the case under consideration and are summarized in Tab.~\ref{tab:boundaryconditions} of the appendix. 
Fig.~\ref{fig:phasediagramaxi} shows the resulting phase diagram together with a comparison of bending energies. 
The dark green line \fbox{$\mathrm{C}_1$} in the diagram separates the region of contact-circle solutions from the region of contact-area solutions. 
For $v_r<0.65$ these solutions coexist with the free sickle-shaped tori. Comparing the respective bending energies one finds a discontinous shape transition (lime green line with long dashes \fbox{$\mathrm{D}_1$} in the phase diagram), below which the free sickle-shaped tori are the global minima. A first estimate of the upper boundary of the region of confined axisymmetric tori (short-dashed lime green curve \fbox{$\mathrm{D}_2^p$}) can be found by comparing the bending energies of these solutions with those of the conformal transformations of the free tori. We would like to stress that these free shapes are \textit{not} equilibrium solutions in general but serve as a means to determine the regions of the phase diagram in which non-axisymmetric shapes with a lower bending energy than the axisymmetric solutions exist. We will see below that the curve including all solutions with and without axisymmetry (black curve \fbox{$\mathrm{D}_2$} in Fig.~\ref{fig:phasediagramfinal}, see next section) lies below the boundary discussed here. 

Consider first the free sickle-shaped torus with a reduced volume of $v_r=0.65$ (lime green square). It has the smallest bending energy of the whole branch \fbox{1} of sickle-shaped tori (see Fig.~\ref{fig:phasediagramaxi_b}). A conformal transformation of this shape along the $\VECx$ axis will not change the energy but increase the reduced volume (black dashed-dotted line \fbox{T}). With a rescaling of such a shape or a subsequent conformal transformation along $\VECz$ (which both do not change the energy as well) every point of the phase diagram left of the curve \fbox{T} can be reached. The resulting surface has to be compared to the other equilibrium shapes of same $(a,v_r)$. Along the black dashed-dotted line one finds that the energy of the conformally transformed free sickle-shaped torus is lower than the energy of the axisymmetric contact-area solution for $v_r \ge 0.76$ (black square). Repeating this comparison for smaller area, one obtains the short-dashed lime green curve \fbox{$\mathrm{D}_2^p$} along which the energy of the conformally transformed free sickle-shaped torus equals that of the corresponding axisymmetric contact-area/contact-line solution. The curve ends in the point at which it intersects with the branch \fbox{2} of the free solutions (lime green circle). The part of the curve  \fbox{$\mathrm{D}_2^p$} to the right of the black square can be found by comparing the conformal transformations of all other free tori of the sickle-shaped branch \fbox{1} with the axisymmetric contact-area solutions. 
Both curves, \fbox{$\mathrm{D}_2^p$} and \fbox{1}, end in the point $(a,v_r)=(2,0)$, which corresponds to a surface consisting of two connected identical spheres with an infinitesimal hole in each of the poles (see Sec.~\ref{sec:freesolutions}). The short-dashed black curve \fbox{L} in Fig.~\ref{fig:phasediagramaxi_a} indicates the conformal transformations of the limit shape along $\VECx$ and delimits the regions of shapes with and without self-contact (see next section). The analytical expression for this curve is given by:
\begin{equation}
  v_r = \frac{1-(a-1)^\frac{3}{2}}{a^\frac{3}{2}}
  \; .
 \label{eq:limitselfcontacts}
\end{equation}
To understand the region of confined tori in the center of the phase diagram, consider curves of constant area $a$ (curves \fbox{I} - \fbox{III} in Fig.~\ref{fig:phasediagramaxi}). For a small reduced volume $v_r$ the solutions are free sickle-shaped tori which can be mapped on one another \textit{via} a simple rescaling. Their bending energy does not depend on $a$ and corresponds to the curve \fbox{1} in Fig.~\ref{fig:phasediagramaxi_b}. When we cross the curve \fbox{$\mathrm{D}_1$} in the phase diagram (yellow, purple, and pink squares, respectively), a discontinous shape transition to a confined state is energetically favorable. Depending on the value of $a$ the membrane can either adopt a contact-circle (curve \fbox{I}) or a contact-area (curves \fbox{II} and \fbox{III}) solution. 
In the former case ($a=0.7$), the solution adopts a free shape again above $v_r=0.63$ (yellow circle). The corresponding bending energy is given by the curves \fbox{2} (between yellow and blue circle) and \fbox{3} (to the right of the blue circle) in Fig.~\ref{fig:phasediagramaxi_b}. In the latter case ($a=0.9$ and $a=1.1$),
we can observe in more detail how the (preliminary) upper limit \fbox{$\mathrm{D}_2^p$} of the region of confined axisymmetric solutions has been obtained. The respective bending energies of the confined solutions (curves \fbox{II} and \fbox{III}) decrease first before increasing up to the point where the energy is equal to one of a conformally transformed free solution (purple and pink circles on \fbox{$\mathrm{D}_2^p$}, respectively). One would thus expect a transition to a non-axisymmetric free torus for higher $v_r$. In reality, however, the membrane will not become a free torus again but prefers to adopt a non-axisymmetric confined shape as we will see when including \textit{all} solutions in our discussion.

%%%%%%%%%%%%%%%%%%%%%%%%%%%%%%%%%%%%%%%%%%%%%%%%%%%%%%%%%%%%%%%%%%%%%%%%%%%%%%%%

\section{Confined non-axisymmetric solutions \label{sec:confinednonaxisymmetricsolutions}}
%%%%%
\begin{figure}[t]
\centering
\begin{minipage}{0.2\textwidth}
\subfigure[][]{\label{fig:comparison_a}\includegraphics[width=\textwidth]{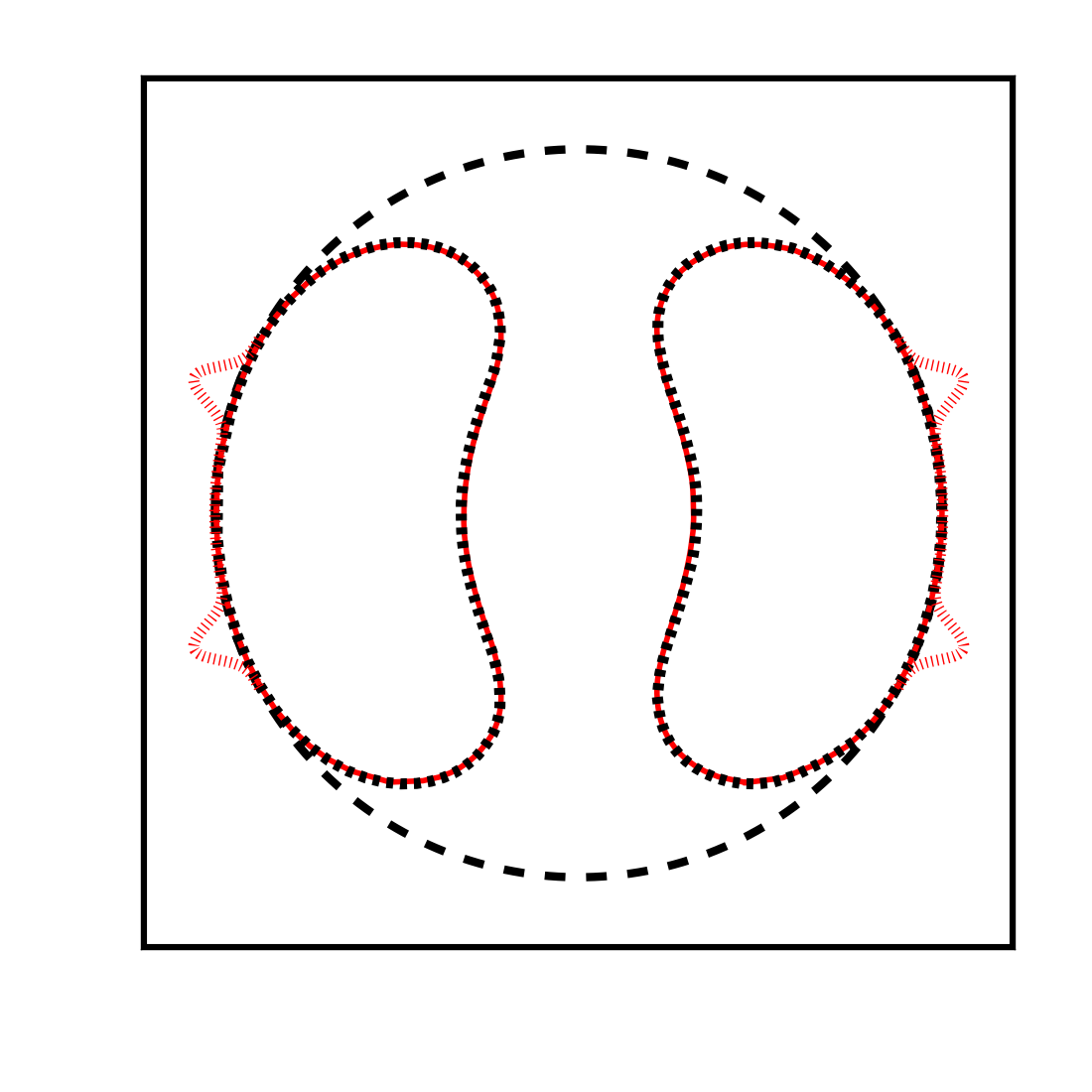}}\\
\subfigure[][]{\label{fig:comparison_b}\includegraphics[width=\textwidth]{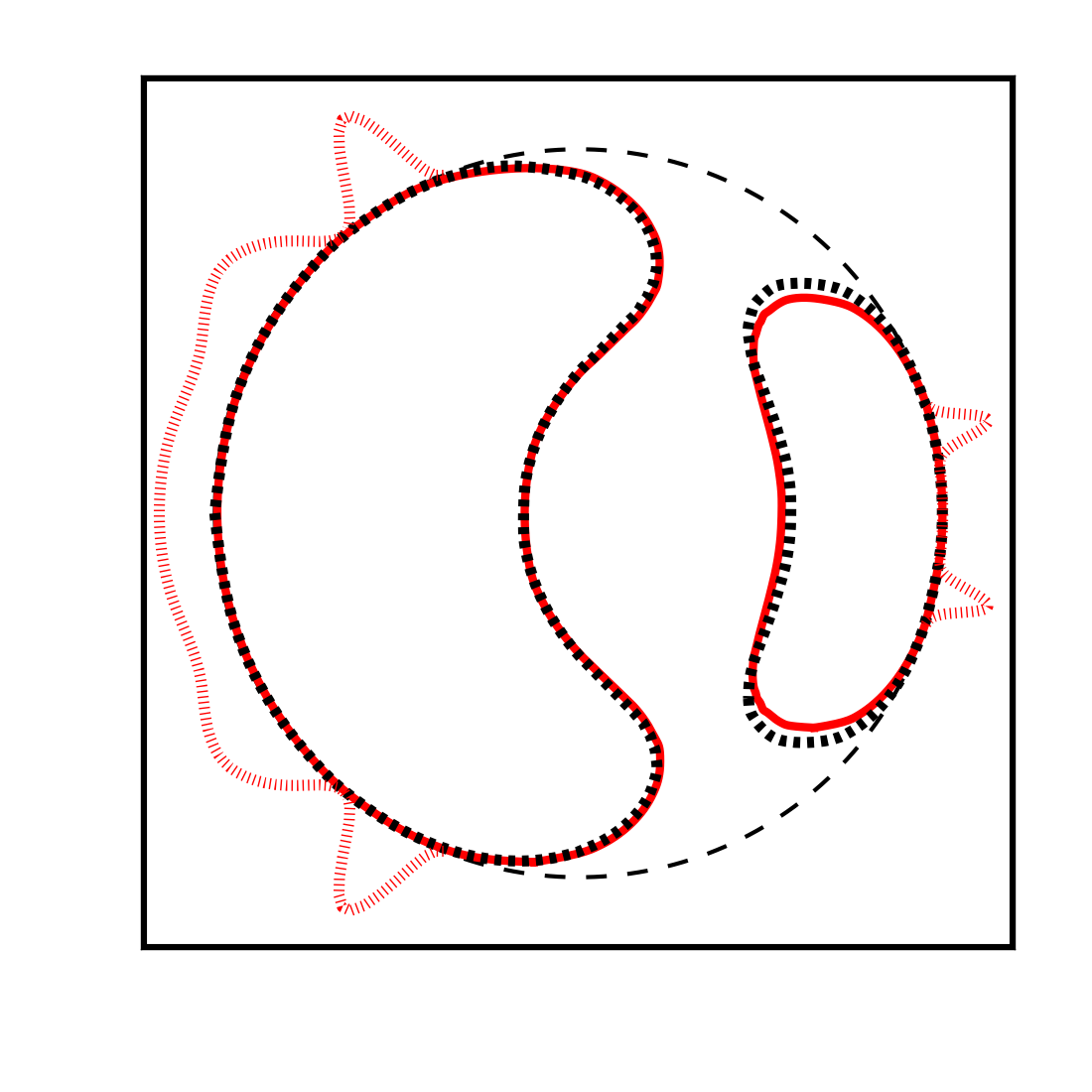}}
\end{minipage}
\hspace*{2cm}
\begin{minipage}{0.45\textwidth}
\subfigure[][]{\label{fig:comparison_c}\includegraphics[width=\textwidth]{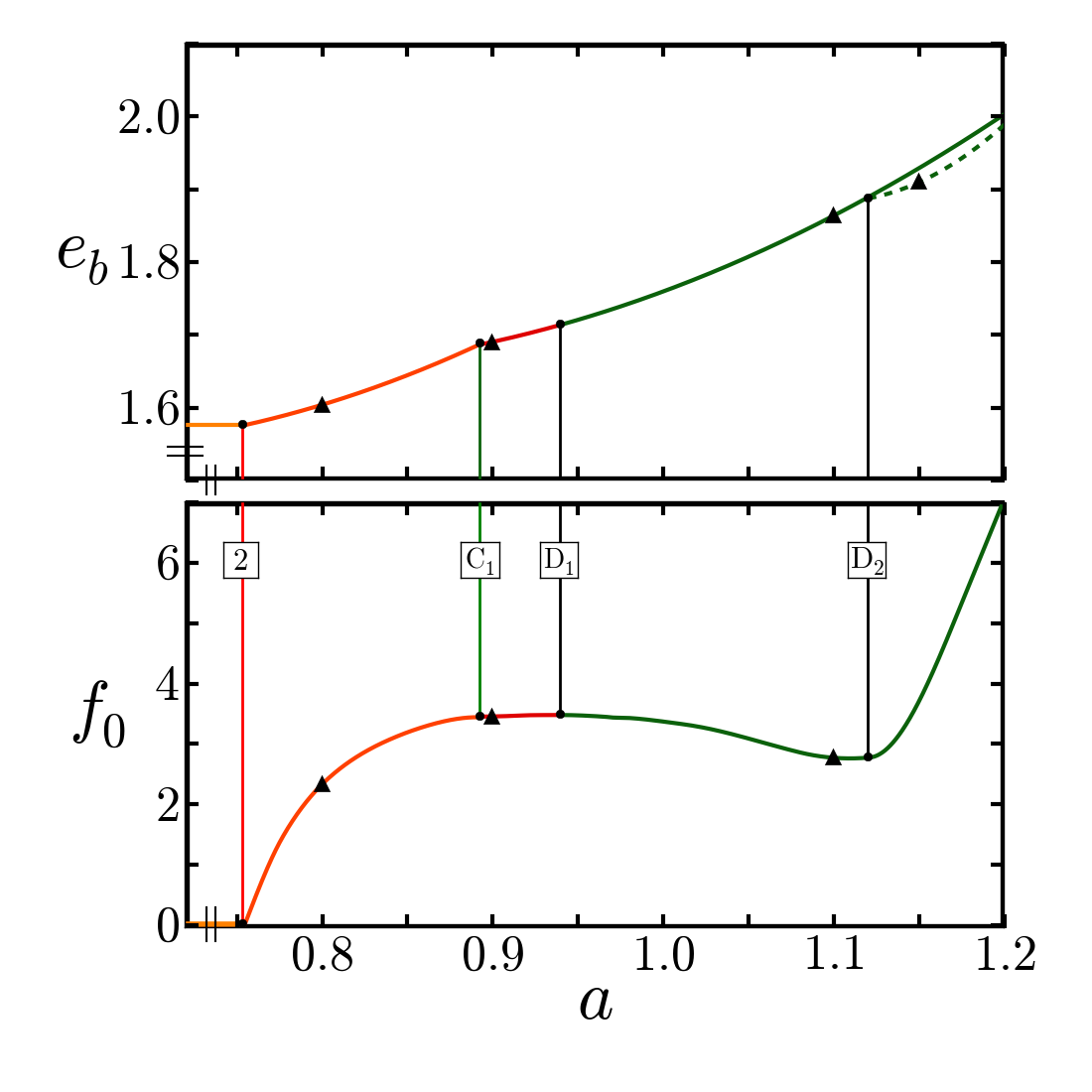}}
\end{minipage}
\caption{(Color online) Comparison between (a)  the numerical Runge-Kutta result (black dotted) and the result of a finite element simulation (red solid) for $(a,v_r)=(1.1,0.7)$, and (b)  the conformal transformation of minimal bending energy (black dotted) and the result of a finite element simulation (red solid) for $(a,v_r)=(1.15,0.7)$ (see Fig.~\ref{fig:profiles} for a three-dimensional representation of the shapes). 
The red dotted curve outside of the container indicates position and relative value of the container force $|\VECf_a^{C}|$ in the finite element simulations. 
(c) Scaled bending energy $e_b$ and container force density $f_0$ at the contact line as a function of $a$ for fixed reduced volume $v_r=0.7$. The green dotted line corresponds to non-axisymmetric contact-area solutions and was obtained with the help of the finite element simulations. All other curves were determined with the Runge-Kutta method. 
Colors and labels are in accordance with the final phase diagram (Fig.~\ref{fig:phasediagramfinal}). The black triangles correspond to the shapes shown in Fig.~\ref{fig:profiles}.}
\label{fig:comparison}%
\end{figure}
%%%%%

As a starting point we consider conformal transformations given by the combination of the equations~(\ref{eq:specialconformaltransformation}) and (\ref{eq:conformaltransformationconfined}) and apply them on the confined axisymmetric solutions $\VECX$ of the previous section:
\begin{equation}
  \bar{\VECX} = \frac{\frac{\VECX}{|\VECX|^2} + \VECLambda}{|\frac{\VECX}{|\VECX|^2} + \VECLambda|^2} (1-\Lambda^2) + \VECLambda
  \; ,  \quad  \text{with } \VECLambda= \textbf{cst} \in\RR^3  \; .
  \label{eq:conformaltransformationconfinedgeneral}
\end{equation}
Every point of the membrane in contact with the spherical confinement is mapped on some point of the sphere again \textit{via}  Eq.~(\ref{eq:conformaltransformationconfinedgeneral}). 
The resulting contact surface/contact line is different from the original spherical segment/circle. However, one can show that the local curvature condition at the boundary/boundaries of the free part of the membrane is still fulfilled (see appendix~\ref{app:conformaltransformations}). 
To be sure that the obtained solutions are in equilibrium, one also has to check whether the total force on the membrane exerted by the container is zero. 
This is the case for the axisymmetric solutions $\VECX$, where the container exerts a constant normal delta force in each point of the contact line(s) to equilibrate the system. The corresponding container force density $f_0$ follows from the boundary conditions of the Runge-Kutta scheme and is given by $f_0=\ddot{\psi}_0$ (see appendix ~\ref{app:RungeKutta}, Eq.~(\ref{eq:stresstensorprojection}), and---for an example---the lower part of Fig.~\ref{fig:comparison_c}). 
A closer look at the conformal transformations $\bar{\VECX}$ of the axisymmetric shapes reveals, however, that the total force on the transformed membrane exerted by the container is not necessarily zero (see appendix~\ref{app:conformaltransformations}). 

In analogy to Ref.~\cite{Juelicher1993B} one might, nevertheless, hope that the resulting surfaces are close to the true equilibrium non-axisymmetric solutions. Assuming this for the moment we can compare bending energies in the same manner as in the previous section. 
For every confined axisymmetric solution one can identify the region in the phase diagram in which its conformal transformations can be found. 
To identify the shape of minimal bending energy for a fixed $(a,v_r)$ in this set of solutions, one thus has to identify all axisymmetric shapes (of different  $(a',v'_r)$) which can be transformed in such a way that the new surface has the 
area and reduced volume $(a,v_r)$. Comparing the energies of all these shapes with the axisymmetric solution of same $(a,v_r)$ allows to determine an approximate boundary \fbox{$\mathrm{D}_2$} which separates the region of axisymmetric and non-axisymmetric global minimum solutions. 

To check how well the obtained solutions approximate the real ones, we have searched for equilibrium shapes with the help of numerical simulations. The respective finite element method has already been used for confined membranes of spherical topology \cite{OsmanNorbertMartin2012A,OsmanNorbertMartin2012B}. In the simulations the membrane is discretized and evolved according to Newton's equations of motion until a balance of forces is reached (see appendix~\ref{app:finiteelementsimulations}). To find an equilibrium surface for a fixed $(a,v_r)$, we have used a discretization of the obtained shape of minimal energy as initial mesh. It turns out that the axisymmetric solutions and the corresponding results of the finite element simulations coincide within the numerical error which is about $10^{-3}$ in scaled units for all parameters (see Fig.~\ref{fig:comparison_a}). The delta forces exerted by the container are smeared out slightly but situated at the contact lines as expected. The necks diffuse away from the axisymmetric ground state very easily for non-vanishing noise during the simulations, similar to what has been observed for unconfined toroidal vesicles before \cite{Sakashita2015}.
The non-axisymmetric solutions, and consequently the phase boundary, are approximated remarkably well by the respective conformal transformations (with an error of about $10^{-2}$ in $e_b$ for shapes close to the phase boundary). In addition to the smeared-out delta forces at the contact lines one now observes additional forces at the contact area (see Fig.~\ref{fig:comparison_b}).

%%%%%
\begin{figure}[t]
\centering
\subfigure[][]{\label{fig:phasediagramfinal_a}\includegraphics[width=0.48\textwidth]{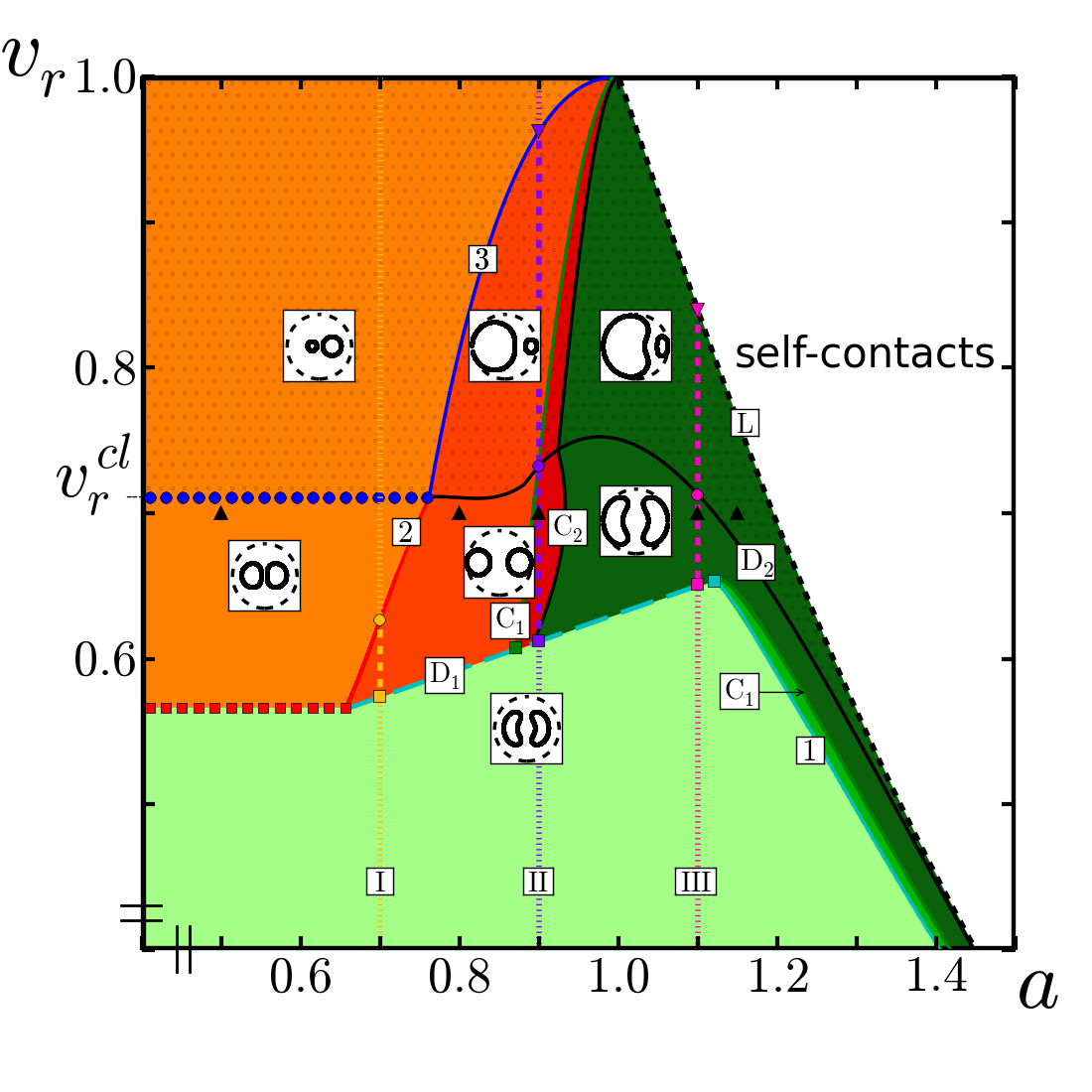}}
\hfill
\subfigure[][]{\label{fig:phasediagramfinal_b}\includegraphics[width=0.48\textwidth]{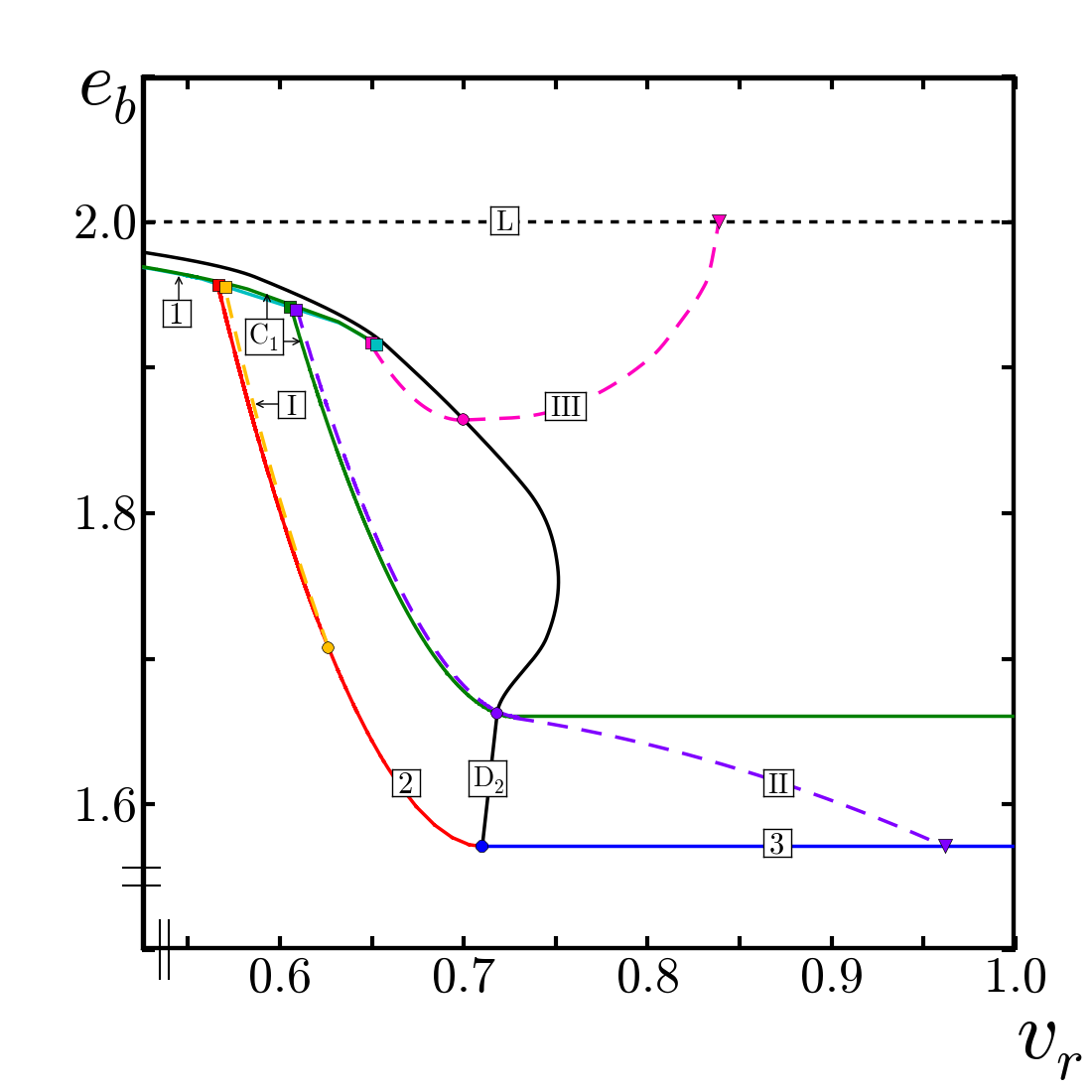}}
\caption{(Color online)  (a) Phase diagram $(a,v_r)$ showing all global minima. Regions of sickle-shaped tori are shown in green, regions of circular tori in orange/red.  Non-axisymmetric regions are hatched. The black triangles correspond to the shapes shown in Fig.~\ref{fig:profiles}. The curve \fbox{$\mathrm{D}_2$} separates the region of confined axisymmetric and confined non-axisymmetric solutions. For a description of the other curves, see Fig.~\ref{fig:phasediagramaxi}.  
(b) Scaled bending energy $e_b$ as a function of $v_r$. This figure is equivalent to Fig.~\ref{fig:phasediagramaxi_b} except that we have included the confined non-axisymmetric solutions as well.}
\label{fig:phasediagramfinal}%
\end{figure}
%%%%%

In summary one obtains Fig.~\ref{fig:phasediagramfinal} which shows all global minima assembled in one figure. 
We can now observe what has been anticipated in the previous section: a membrane of large constant area (curves \fbox{II} and \fbox{III}) breaks axisymmetry but stays in contact with the container when we cross the curve \fbox{$\mathrm{D}_2$} towards higher $v_r$. Depending on the value of $a$ the membrane either adopts the shape of a free conformal transformation of the Clifford torus when crossing the curve \fbox{3} (purple triangle) or tends towards the limit torus, which consists of two identical spheres connected \textit{via} two infinitesimal holes (pink triangle on curve \fbox{L}).  

It is instructive to consider the case of constant reduced volume as well. The upper part of Fig.~\ref{fig:comparison_c} shows the scaled bending energy as a function of membrane area for a reduced volume of $v_r=0.7$ (compare Fig.~\ref{fig:profiles} in Sec.~\ref{sec:confinedaxisymmetricsolutions}). For small area the membrane is not in contact with the confining sphere and the bending energy does not depend on $a$. When increasing $a$ contact emerges and the energy starts increasing monotonically. We first find contact-circle, then contact-area solutions. At $a=1.12$ the membrane can break axisymmetry to lower its energy (dotted green line in Fig.~\ref{fig:comparison_c}) and adopts a confined non-axisymmetric shape. 

For larger values of fixed reduced volume $v_r$ a similar evolution is observed: for small area $a$ the solution corresponds to a conformal transformation of the Clifford torus (hatched orange region). By increasing $a$ we cross the blue curve defined by Eqs.~(\ref{eq:touchingtransformedCliffordtorus}). The non-axisymmetric solutions in this region are in contact with the container in one horizontal circle (hatched light red region). When crossing the left black solid line the contact line extends to a zone of contact. The corresponding non-axisymmetric tori are first circular (hatched dark red region) and, upon increasing $a$ even further, sickle-shaped (hatched dark green region). All the shapes lying to the right of the curve~(\ref{eq:limitselfcontacts}) exhibit self-contacts and are not studied in the present paper.

%%%%%%%%%%%%%%%%%%%%%%%%%%%%%%%%%%%%%%%%%%%%%%%%%%%%%%%%%%%%%%%%%%%%%%%%%%%%%%%%

\section{Comparison to spherical topology \label{sec:comparisonsphericaltopology}}
%%%%%%%%%
\begin{table}
\begin{tabular}{|c||c|c|c|}
\hline
$(a,v)$ & $ e_\kappa^\text{sph}$  & $ e_\kappa^\text{tor}$ & toroidal membrane shape \\
\hline
$(1.1,0.9)$ & 1.91 & 1.88 & non-axi. confined \\ \hline
$(1.2,0.9)$ & 1.97 & 1.97 & non-axi. confined \\ \hline
\hline
$(1.1,0.8)$ & 1.95 & 1.87 & axi. confined \\ \hline
$(1.2,0.8)$ & 1.98 & 1.94 & axi. confined\\ \hline
$(1.3,0.8)$ & 1.99 & 1.97 & non-axi. confined \\ \hline
\hline
$(1.1,0.7)$ & 1.99 & 1.94 & not confined \\ \hline
$(1.2,0.7)$ & 1.99 & 1.97 & not confined\\ \hline
$(1.3,0.7)$ & 2.00 & 1.98 & not confined \\ \hline
$(1.4,0.7)$ & 2.00 & 1.98 & axi. confined\\ 
\hline
\end{tabular}
\caption{Comparison of energies $e_\kappa$ of a vesicle of spherical (from Ref.~\cite{OsmanNorbertMartin2012A}) and toroidal topology as a function of scaled area $a$ and volume $v$.}
\label{tab:bendingenergycomparison}
\end{table}
%%%%%%%%%

Knowing the global minima of the confined toroidal membrane vesicles we are now in a position to compare its bending energy~(\ref{eq:bendingenergy}) to that of a confined spherical membrane of same area and volume. For a better comparison we again scale the energy with the constant $8\pi\kappa$, $i.e.$, the energy $E_\kappa$ of a perfect spherical vesicle: $e_b=E_b/(8\pi\kappa) = e_\kappa+e_{\bar{\kappa}}$.

For the \textit{free} membrane vesicles one finds that the toroidal ground state has a lower energy $e_\kappa$ than the corresponding spherical one as long as $0<v_r<0.72$ (see Ref.~\cite{Seifert1991A}). For larger values of $v_r$ the opposite is true. 
The second term in the energy, $e_{\bar{\kappa}}$, equals $\frac{1}{2}\frac{\bar{\kappa}}{\kappa}$  for spherical vesicles, whereas it is zero for vesicles of toroidal topology (see Sec.~\ref{sec:model}). Absolute value and sign of $\frac{\bar{\kappa}}{\kappa}$ will decide which topology is preferred energetically. Taking a typical value for fluid lipid mono- and bilayers $\frac{\bar{\kappa}}{\kappa}\approx -1$ \cite{Hu2012} the systems increases its energy by $\Delta e_{\bar{\kappa}}\approx\frac{1}{2}$ when passing from a spherical to a toroidal topology. Since this energy is not compensated by the change $\Delta e_\kappa$ (compare inset of Fig.~\ref{fig:phasediagramfreesolutions} with Fig.~8 of Ref.~\cite{Berndl91}) we find that the spherical topology is always preferred for the free system.

A similar comparison can be drawn for the confined vesicles. We focus on the case $a\ge 1$ for which we have found axisymmetric solutions of spherical topology \cite{OsmanNorbertMartin2012A,OsmanNorbertMartin2012B} such as the one shown in  Fig.~\ref{fig:topologychange} (\textit{bottom left}). Table~\ref{tab:bendingenergycomparison} depicts the values of $e_\kappa$ for both topologies. For a better comparison with the results of Refs.~\cite{OsmanNorbertMartin2012A,OsmanNorbertMartin2012B} we have fixed $v$ instead of $v_r$. One observes that the system decreases its energy $e_\kappa$ when changing its topology from spherical to toroidal. However, as is the case of the free system, this change does not compensate the increase of $\approx \frac{1}{2}$ in the Gaussian term of the energy. Note that the situation changes dramatically if we allow for other values of $\bar{\kappa}$. For positive $\bar{\kappa}$ the toroidal topology is favored energetically implying the possibility of a topological transition without any loss of energy.

%%%%%%%%%%%%%%%%%%%%%%%%%%%%%%%%%%%%%%%%%%%%%%%%%%%%%%%%%%%%%%%%%%%%%%%%%%%%%%%%

\section{Conclusion \label{sec:conclusion}}
In this article we have studied the morphology of toroidal fluid membrane vesicles in spherical confinement. A combination of analytical theory, numerical calculations and finite element simulations allowed us to determine the ground states as a function of area and volume of the vesicle. 
In contrast to confined membrane vesicles of spherical topology we have found non-axisymmetric shapes which do not exhibit any self-contacts. 
A comparison of energies revealed, that the spherical topology is preferred for typical values of the material parameters. 

However, this study was performed assuming zero temperature. To be closer to the real biological world, one would have to include the effects of thermal fluctuations. In extensive bilayer fluid membrane phases they can significantly affect the competition between spherical morphologies and surfaces of higher genus \cite{Golubovic94,Morse94,Gompper00}. In these cases an attractive interaction appears in addition to Helfrich's steric repulsion \cite{Helfrichrepulsion}. Guided by experiments on real mitochondria \cite{john2005}, we expect a similar competition in a biologically significant setting, whenever the membrane is in self-contact. This gives a hint to what can happen in the region of self-contacts in the final phase diagram (Fig. 6): the toroidal vesicle touches itself, inducing eventually a topology change towards a higher genus (``a sphere with two handles''). 

A study of these shapes and the effects of the temperature goes beyond the scope of this paper but remains an interesting open problem for the future. The present model can as well be extended to incorporate a constant adhesion energy density due to the contact between membrane and confinement.

%%%%%%%%%%%%%%%%%%%%%%%%%%%%%%%%%%%%%%%%%%%%%%%%%%%%%%%%%%%%%%%%%%%%%%%%%%%%%%%%

\begin{acknowledgments}
The authors thank Osman Kahraman and Norbert Stoop for helpful discussions and Etienne Gallant for his support. The PMMS (P\^ole Messin de Mod\'elisation et de Simulation) is acknowledged for providing the computer time. 
\end{acknowledgments}

%%%%%%%%%%%%%%%%%%%%%%%%%%%%%%%%%%%%%%%%%%%%%%%%%%%%%%%%%%%%%%%%%%%%%%%%%%%%%%%%
\appendix

\section{Numerical method for the axisymmetric case \label{app:RungeKutta}}
%%%%%
\begin{figure}[t]
\centering
\includegraphics*[width=0.35\textwidth,bb= 160 380 450 680]{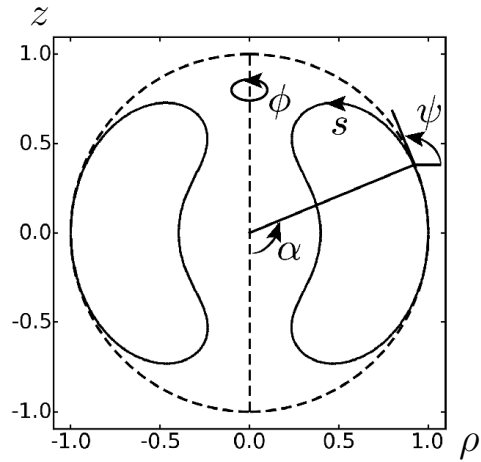}
\caption{Parametrization of an axisymmetric toroidal membrane vesicle in spherical confinement. The membrane consists of a spherical segment in contact with the container and a free part which is given by the solution of the Hamilton equations~(\ref{eq:Hamiltonequations}) together with the appropriate boundary conditions~(Eqs.~(\ref{eq:boundaryconditions})/Tab.~\ref{tab:boundaryconditions}). }%
\label{fig:system}%
\end{figure}
%%%%%
The details of the numerical method we have used to find axisymmetric equilibrium solutions can be found in appendix~A of Ref.~\cite{OsmanNorbertMartin2012B} and the references cited therein. The confined membrane consists of two parts, one which is in contact with the spherical container of unit size and one which bulges inside to minimize its elastic energy. We parametrize the shape of the membrane in terms of the angle-arc length parametrization $\psi(s)$ (see Fig.~\ref{fig:system}), where $\psi$ is the angle between the horizontal axis and the tangent of the membrane.  The equation describing the evolution of $\psi(s)$ along the cross section of the free part is an ordinary differential equation of fourth order which can be rewritten with the help of Hamilton's formalism.

The scaled energy of the free membrane is given by:
\begin{equation}
  \tilde{E} = E/(\pi\kappa) = \int_{\underline{s}}^{\bar{s}} \romd s \; \tilde{L}
    = \int_{\underline{s}}^{\bar{s}} \romd s \;  \left[\rho \left( \dot{\psi} + \frac{\sin{\psi}}{\rho} \right)^2 + 2\tilde{\sigma} \rho  + \tilde{P} \rho^2 \sin{\psi} + \lambda_\rho (\dot{\rho} - \cos{\psi}) + \lambda_z (\dot{z}-\sin{\psi})  \right]
    \; ,
\end{equation}
where $\tilde{\sigma}=\sigma/\kappa$ and $\tilde{P}=P/\kappa$ are the scaled surface tension and pressure difference, respectively. For the Clifford torus one finds $\tilde{\sigma}=\tilde{P}=0$. The Lagrange multiplier functions $\lambda_\rho$ and $\lambda_z$ couple the angle-arc length parametrization to the Euclidean coordinate system $(\rho,z)$. 
The conjugate momenta $p_i=\frac{\partial \tilde{L}}{\partial \dot{q_i}}$ of the system are:
$p_\psi = 2\rho(\dot{\psi}+\frac{\sin{\psi}}{\rho})$, $p_\rho = \lambda_\rho$, and $p_z=\lambda_z$. In contrast to the case considered in Ref.~\cite{OsmanNorbertMartin2012B} the momentum $p_z$ does not vanish \cite{Hamiltonian94}. 
The Hamiltonian is thus:
\begin{equation}
  \tilde{H} = \sum_i \dot{q}_i p_i - \tilde{L}= \frac{p_\psi^2}{4\rho} - \frac{p_\psi \sin{\psi}}{\rho} + p_\rho \cos{\psi} + p_z \sin{\psi} - 2 \tilde{\sigma} \rho - \tilde{P} \rho^2 \sin{\psi}
  \; ,
\end{equation}
yielding the following shape equations:
\begin{subequations}
\label{eq:Hamiltonequations}
\begin{eqnarray}
  \dot{\psi} & = & \frac{p_\psi}{2\rho} - \frac{\sin{\psi}}{\rho} \; , \label{eq:Hamiltonequation1}\\
  \dot{\rho} & = & \cos{\psi}  \; ,\label{eq:Hamiltonequation2}\\
  \dot{z} & = & \sin{\psi}  \; ,\label{eq:Hamiltonequation3} \\
  \dot{p}_\psi & = & \left( \frac{p_\psi}{\rho} + \tilde{P} \rho^2 - p_z \right) \, \cos{\psi} + p_\rho \sin{\psi} \; , \label{eq:Hamiltonequation4}\\
  \dot{p}_\rho & = & \frac{p_\psi}{\rho} \left(  \frac{p_\psi}{4 \rho} - \frac{\sin{\psi}}{\rho}  \right) + 2 \tilde{\sigma} + 2 \tilde{P} \rho \sin{\psi} \; , \label{eq:Hamiltonequation5}\\
  \dot{p}_z & = & 0 \label{eq:Hamiltonequation6}\; ,
\end{eqnarray}%
\end{subequations}%
which are solved using a standard shooting method \cite{NumRec} subject to the boundary conditions:
\begin{subequations}
  \label{eq:boundaryconditions}
\begin{eqnarray}
  \tilde{H} & = & 0 \; , \label{eq:Hzero}\\
  \rho(\underline{s}) & = & \rho_0 \, , \quad z(\underline{s})  = z_0 \; , \\
  \psi(\underline{s}) & = & \alpha \; .
\end{eqnarray}%
\end{subequations}%
%
%%%%%%%%%
\begin{table}
\begin{tabular}{|c|c||c|c|c|c|c|c|c|c|}
\hline
& Section & $\rho_0$ & $z_0$ & $\alpha$ & $\dot{\psi}_0$ & $\ddot{\psi}_0 $ & $p_z$ & $p_\rho$ & $p_\psi$ \\
\hline
Free solutions ($xy$ sym.)  & \ref{sec:freesolutions} & 1 & 0  & $\frac{\pi}{2}$ & \textit{bisection} & 0 & $-\dot{\psi}_0^2 + 1 +2\tilde{\sigma} + \tilde{P}$ & 0 & $2 (\dot{\psi}_0 + 1)$ \\
Free toroidal stomatocytes & \ref{sec:freesolutions}&  1 & 0  & $\frac{\pi}{2}$ & \textit{bisection} & \textit{scan} &  $-\dot{\psi}_0^2 + 1 +2\tilde{\sigma} + \tilde{P}$ & $2\ddot{\psi}_0$ & $2(\dot{\psi}_0 + 1)$  \\ 
Circular contact line & \ref{sec:confinedaxisymmetricsolutions}& 1 & 0 & $\frac{\pi}{2}$ & \textit{bisection} & \textit{scan} & '' &  '' & ''  \\
Transition contact line $\to$ area & \ref{sec:confinedaxisymmetricsolutions}& 1 & 0 & $\frac{\pi}{2}$ & $1$ & \textit{bisection} & '' &  '' & ''  \\
Contact area & \ref{sec:confinedaxisymmetricsolutions} & $\sin{\alpha}$ & $-\cos{\alpha}$ & \textit{bisection} & $1$ & \textit{see text} & \textit{scan} &  \textit{via Eq.}~(\ref{eq:Hzero}) & $4 \sin{\alpha}$  \\
\hline
\end{tabular}
\caption{Boundary conditions for the different cases discussed in this article.}
\label{tab:boundaryconditions}
\end{table}
%%%%%%%%%
The values $\rho_0$, $z_0$, and $\alpha$ depend on the case under consideration (see Tab.~\ref{tab:boundaryconditions}). It is sufficient to scan $\tilde{\sigma}$ and $\tilde{P}$ to obtain the free solutions with $xy$ symmetry as well as the transition line of shapes where the contact line extends to a zone of contact.  For the other cases a third parameter has to be scanned as well. In the case of the free toroidal stomatocytes  and the contact-circle solutions we have fixed the initial value of $\ddot{\psi}$. For the contact-area solutions it was more practical to scan $p_z$ instead. The value of $\ddot{\psi}_0$ can then be obtained by deriving Eq.~(\ref{eq:Hamiltonequation1}), equating it with Eq.~(\ref{eq:Hamiltonequation4}) and inserting the values of the other parameters in the resulting equation.

In all cases the values of $\psi$, $\rho$, $z$,  and the corresponding conjugate momenta at $s=\underline{s}$ are determined by the boundary conditions for given $\alpha$, $\dot{\psi}_0$, and $\ddot{\psi}_0$. The Hamilton equations~(\ref{eq:Hamiltonequations}) can then be integrated with a fourth-order Runge-Kutta method. The integration is stopped when the curve returns to the spherical confinement. A bisection on either $\alpha$, $\dot{\psi}_0$ or $\ddot{\psi}_0$ (depending on the case, see Tab.~\ref{tab:boundaryconditions} again) determines the correct profiles for which $\psi(\bar{s})=\frac{5\pi}{2}-\alpha$. 

Note that we have to fix pressure and surface tension for this method. Area and volume can only be calculated for a known profile \textit{a posteriori}. The equations  $\dot{A}=2\pi\rho$ and $\dot{V}=\pi\rho^2\sin{\psi}$ have to be integrated from $s=\underline{s}$ to $\bar{s}$ to find the contribution to area and volume of the free part of the membrane. Adding the result to the contribution
of the part in contact with the container, $A_c=-4\pi \cos{\alpha}$ and $V_c=\frac{2}{3} \pi (-3\cos{\alpha} + \cos^3{\alpha})$, finally yields the values for the whole membrane. These allow to assemble the obtained shapes in the geometrical phase diagrams (Figs.~\ref{fig:phasediagramfreesolutions}, \ref{fig:phasediagramaxi}, and \ref{fig:phasediagramfinal}). The Runge-Kutta integration is performed with an integration step-size $h=10^{-4}$ in reduced units implying a numerical error of $10^{-3}$ of the values of $a$ and $v_r$.

%%%%%%%%%%%%%%%%%%%%%%%%%%%%%%%%%%%%%%%%%%%%%%%%%%%%%%%%%%%%%%%%%%%%%%%%%%%%%%%%

\section{Conformal transformations of confined axisymmetric solutions \label{app:conformaltransformations}}

We consider the conformal transformation of an axisymmetric surface $\VECX$ given by Eq.~(\ref{eq:conformaltransformationconfinedgeneral}):
\begin{equation*}
  \bar{\VECX} = \frac{\frac{\VECX}{|\VECX|^2} + \VECLambda}{|\frac{\VECX}{|\VECX|^2} + \VECLambda|^2} (1-\Lambda^2) + \VECLambda
  \; ,  \quad  \text{with } \VECLambda= \textbf{cst} \in\RR^3  \; , 
\end{equation*}
yielding a deformed surface $\bar{\VECX}$ which is only axisymmetric when $\VECLambda$ is parallel to the symmetry axis.
The tangent vectors of the transformed surface are given by (see chapter 20 of Ref.~\cite{Gray2006})
\begin{equation}
  \bar{\VECe}_a = \partial_a \bar{\VECX} = \frac{(1-\Lambda^2)}{|\VECX_0+\VECLambda|^2} \left( \VECe_a^0 
   -   \frac{2 [(\VECX_0 +\VECLambda) \cdot \VECe_a^0 ] (\VECX_0+\VECLambda)}{|\VECX_0+\VECLambda|^2} \right)
\end{equation}
with $\VECX_0=\frac{\VECX}{\VECX^2}$ and $\VECe_a^0=\partial_a \VECX_0 = (\VECe_a - 2 (\VECX\cdot\VECe_a) \VECX)/|\VECX|^2$. The metric of the transformed surface follows as
\begin{equation}
  \bar{g}_{ab} = \frac{(1-\Lambda^2)^2}{|\VECX_0+\VECLambda|^4} \; g^0_{ab} =  \frac{(1-\Lambda^2)^2}{|\frac{\VECX}{|\VECX|^2} +\VECLambda|^4\, |\VECX|^4} \; g_{ab}
  \; .
\end{equation}
We orient the normal towards the outside of the vesicle. The transformed normal is then given by:
\begin{equation}
  \bar{\VECn} = -\VECn_0 +  \frac{2 [\VECn_0 \cdot (\VECX_0 +\VECLambda) ] (\VECX_0+\VECLambda)}{|\VECX_0+\VECLambda|^2}
   \; ,
\end{equation}
where $\VECn_0 = -\VECn + \frac{2 (\VECn\cdot\VECX)\VECX }{|\VECX|^2}$. The extrinsic curvature tensor $\bar{K}_{ab}=-\bar{\VECn}\cdot\partial_a\bar{\VECe}_b$ can be obtained by calculating $\partial_a\bar{\VECe}_b$
and projecting it onto the normal
\begin{equation}
    \bar{K}_{ab} =   \frac{(1-\Lambda^2)}{|\VECX_0+\VECLambda|^2} \left\{ -K_{ab}^0 + \frac{2 g_{ab}^0 [\VECn_0 \cdot (\VECX_0 +\VECLambda)]}{|\VECX_0+\VECLambda|^2}
    \right\} 
    \; .
\end{equation}
When the coordinate lines on the original surface before the transformation coincide with the lines of curvature, first and second fundamental form are diagonal. These lines will be lines of curvature on the transformed surface as well and the principal curvatures follow as ($i\in\{1,2\}$):
\begin{equation}
  \bar{K}_i = \frac{\bar{K}_{ii}}{\bar{g}_{ii}} = \frac{|\VECX_0+\VECLambda|^2}{(1-\Lambda^2)} \left\{ -K_{i}^0 + \frac{2  [\VECn_0 \cdot (\VECX_0 +\VECLambda)]}{|\VECX_0+\VECLambda|^2} \right\}
  \; .
  \label{eq:principalcurvaturesconformal}
\end{equation}
where $K_i^0=- |\VECX|^2 K_{i} + 2\,\VECn\cdot\VECX$. The derivative is given by
\begin{equation}
  \partial_a\bar{K}_i  = \frac{1}{(1-\Lambda^2)} \left\{ - |\VECX_0+\VECLambda|^2 \partial_a K_{i}^0 
  + 2 (\VECX_0+\VECLambda) \cdot (  K_{a}^{b\,0} \, \VECe_b^0 - K_{i}^0 \, \VECe_a^0) \right\} 
  \; ,
  \label{eq:derivativeprincipalcurvaturesconformal}
\end{equation}
where we have used the Weingarten equations in the second step. 

To describe the original axisymmetric surface, we choose the coordinates $(s,\varphi)$ . The coordinate lines are lines of curvature with principal curvatures $K_1=K_\perp=\dot{\psi}$ and $K_2=K_\|=\sin{\psi}/\rho$. For each portion of the surface which is in contact with the confining sphere (including the circular contact line(s)) one has $K_\perp=K_\|=1$. Following Eq.~(\ref{eq:principalcurvaturesconformal}) the transformed principal curvatures at the contact lines can be obtained very easily. On the sphere one has $|\VECX|^2=1$, $\VECn=\VECX$ and thus  $K_i^0=1$. 
Moreover, $\VECX_0=\VECX$, $\VECn_0=\VECn$, and one finds that the local contact curvature condition at the contact line is also fulfilled for the transformed shape:
\begin{equation}
  \bar{K}_i =  \frac{1}{(1-\Lambda^2)} \left[ - |\VECX+\VECLambda|^2 + 2  \VECn \cdot (\VECX +\VECLambda)\right]
    =  \frac{1}{(1-\Lambda^2)}  \left[ - 1 - 2 \VECX \cdot \VECLambda - \Lambda^2 + 2   (1 + \VECn\cdot \VECLambda)\right] =1
  \; .
\end{equation}
The tangent and normal vectors on the contact line are given by $\VECe_a^0=\VECe_a$, $\VECn^0=\VECn$,  and
 \begin{equation}
  \bar{\VECe}_a = \frac{(1-\Lambda^2)}{|\VECn+\VECLambda|^2} \left( \VECe_a 
   -   \frac{2 (\VECLambda \cdot \VECe_a) (\VECn+\VECLambda)}{|\VECn+\VECLambda|^2} \right)
\qquad 
  \bar{\VECn} = -\VECn +  \frac{2 (1 + \VECn \cdot \VECLambda) (\VECn +\VECLambda)}{|\VECn +\VECLambda|^2}
  \; .
\end{equation}
The metric is diagonal with $\bar{g}_{ab}=\frac{(1-\Lambda^2)^2}{|\VECn +\VECLambda|^4} g_{ab}$. 
Taking into account that $K_a^{b}$ as well as $K_a^{b\, 0}$ are diagonal, $K_{a}^{b\,0} \, \VECe_b^0 - K_{i}^0 \, \VECe_a^0 
= (K_{a}^0 - K_{i}^0)\VECe_a = 0$. Therefore, the last term in Eq.~(\ref{eq:derivativeprincipalcurvaturesconformal}) vanishes, and one obtains for the derivative of the principal curvatures
\begin{equation}
  \partial_a\bar{K}_i = \frac{- |\VECn+\VECLambda|^2 \partial_a K_{i}^0}{(1-\Lambda^2)} 
   =  \frac{|\VECn+\VECLambda|^2}{(1-\Lambda^2)} \, \partial_a K_{i}
   \; .
\end{equation}
To determine the external forces due to the confinement, consider the stress tensor of the fluid membrane (normalized by $\kappa$) \cite{Guven2006}
\begin{equation}
  \VECf^a = \left\{ K (K^{ab} - \frac{1}{2}K g^{ab}) - \left[\frac{\tilde{P}}{2}(\VECX\cdot\VECn) + \tilde{\sigma}\right] g^{ab} \right\} \VECe_b - \left[\nabla^a K - \frac{\tilde{P}}{2}(\VECX\cdot\VECe^a) \right] \VECn 
  \; .
\end{equation}
Projection onto the unit vector $\VECl = l_a \VECe^a$ perpendicular to the contact line yields the local force density:
\begin{eqnarray}
  l_a \VECf^a & = & \left\{ \frac{1}{2}(K_\perp^2 - K_\|^2) - \left[\frac{\tilde{P}}{2}(\VECX\cdot\VECn) + \tilde{\sigma}\right]\right\}  \VECl- \left[\nabla_\perp K - \frac{\tilde{P}}{2}(\VECX\cdot\VECl) \right] \VECn \nonumber \\
 & = & \left[ \frac{1}{2}\left( \dot{\psi}^2 -\frac{\sin^2{\psi}}{\rho^2}\right) - \left(\frac{\tilde{P}}{2}(\VECX\cdot\VECn) + \tilde{\sigma}\right)\right] \VECe_s 
   - \left[\ddot{\psi} +  \frac{\dot{\psi}\cos{\psi}}{\rho} - \frac{\cos{\psi}\sin{\psi}}{\rho^2} -  \frac{\tilde{P}}{2}(\VECX\cdot\VECl) \right] \VECn
  \; ,
  \label{eq:stresstensorprojection}
\end{eqnarray}
where the second line only holds for an axisymmetric surface. At the circular contact line:
$l_a \VECf^a =  - (\frac{\tilde{P}}{2} + \tilde{\sigma})\,\VECe_s - \ddot{\psi} \,\VECn$.
For the conformally transformed surface one obtains (see Refs.~\cite{Guven2007A,Guven2007B,Guven2013B} for further examples of how stresses change under conformal transformations):
\begin{eqnarray}
   \bar{l}_a \bar{\VECf}^a  & = & -\left(\frac{\tilde{P}}{2} + \tilde{\sigma}\right)\,\left(\frac{ \VECe_s (1 + \Lambda^2) - 2\VECLambda  (\VECLambda \cdot \VECe_s) + 2 \VECLambda \times \VECe_\varphi)}{|\VECn+\VECLambda|^2} \right)  + \ddot{\psi} \,  \left[\frac{-\VECn (1-\Lambda^2) - 2 \VECLambda (1 + \VECn \cdot \VECLambda)}{(1-\Lambda^2)}\right]
   \; .
\end{eqnarray}

To analyze the balance of forces let us consider a conformal transformation along the $\VECz$ axis: $\VECLambda=\Lambda_z\VECz$. The projections of the normal and tangent vectors onto the basis vectors are given by
\begin{eqnarray}
  \VECrho \cdot \VECe_s & = & -\VECz \cdot \VECn = \cos{\psi} \\
  \VECrho \cdot \VECn & = & \VECz \cdot \VECe_s = \sin{\psi}
  \; .
\end{eqnarray}
and thus, with  $\VECn +\VECLambda = \sin{\psi}\VECrho + (\Lambda_z - \cos{\psi}) \VECz$ one obtains the projections
\begin{eqnarray}
  \VECrho \cdot \bar{l}_a \bar{\VECf}^a  & = & -\left(\frac{\tilde{P}}{2} + \tilde{\sigma}\right)\, \left( \frac{\cos{\psi} (1+\Lambda_z^2) - 2 \Lambda_z}{1-2\Lambda_z\cos{\psi}+\Lambda_z^2} \right)  -  \, \ddot{\psi} \, \sin{\psi} \; , \quad \text{and}\\
   \VECz \cdot \bar{l}_a \bar{\VECf}^a   & = &
   -\left(\frac{\tilde{P}}{2} + \tilde{\sigma}\right)\,\left( \frac{\sin{\psi} (1-\Lambda_z^2)}{1-2\Lambda_z\cos{\psi}+\Lambda_z^2} \right)  + \frac{ \ddot{\psi}}{(1-\Lambda_z^2)} \, \left[\cos{\psi}  (1 + \Lambda_z^2) - 2 \Lambda_z \right]
   \; .
\end{eqnarray}
The axisymmetric solutions we consider are symmetric with respect to the $xy$ plane as well. The two circular contact lines at $\psi_+=90^\circ + \gamma$ and $\psi_-=90^\circ - \gamma$ are transformed into two circles of different length. To calculate the total external force let us determine the new positions and lengths of the contact lines. 
Before the transformation we have $\VECX = (\sin{\psi},0,-\cos{\psi})^\romT$. After the transformation:
\begin{equation}
\bar{\VECX} = \lambda \sin{\psi}\, \VECrho + [\lambda (\Lambda_z - \cos{\psi}) + \Lambda_z]\, \VECz
\; ,
\end{equation}
where $\lambda:= (1-\Lambda_z^2)/(1-2\Lambda_z\cos{\psi} + \Lambda_z^2)$. One obtains for the new contact angles:
$\bar{\psi}=90^\circ+\arctan{[\lambda (\Lambda_z - \cos{\psi}) + \Lambda_z)/(\lambda \sin{\psi})]}$. 
The lengths of the contact lines are thus:
$\bar{L} = 2\pi\sin{\bar{\psi}}=2\pi\lambda\sin{\psi}$. 

The horizontal components of the force will equilibrate each other due to the axisymmetry. The vanishing of the force in the vertical direction yields a condition which can be used to find $\Lambda_z$ as a function of the known parameters $\tilde{P}$, $\tilde{\sigma}$, and $\ddot{\psi}$:
\begin{eqnarray}
   0 & = & \VECz \, \cdot (2\pi \lambda_+\sin{\psi_+} (\bar{l}_a \bar{\VECf}^a)_+ + 2\pi  \lambda_-\sin{\psi_-} (\bar{l}_a \bar{\VECf}^a)_-) 
   \nonumber\\
  \Leftrightarrow 0 & = & \left( \frac{- C \cos{\gamma} (1-\Lambda_z^2)^2}{(1+2\Lambda_z\sin{\gamma}+\Lambda_z^2)^2} \right)  + \frac{\left[-\sin{\gamma}  (1 + \Lambda_z^2) - 2 \Lambda_z \right]}{(1+2\Lambda_z\sin{\gamma}+\Lambda_z^2)} +  \left( \frac{- C  \cos{\gamma} (1-\Lambda_z^2)^2}{(1-2\Lambda_z\sin{\gamma}+\Lambda_z^2)^2} \right)  + \frac{ \left[\sin{\gamma}  (1 + \Lambda_z^2) - 2 \Lambda_z \right]}{(1-2\Lambda_z\sin{\gamma}+\Lambda_z^2)} 
  \nonumber
  \; ,
\end{eqnarray}
where we have defined $C:= \frac{\left(\frac{\tilde{P}}{2} + \tilde{\sigma}\right)}{\ddot{\psi}}$.
For $\gamma=0$, $i.e.$, the solutions which are only confined in one circle, this condition simplifies to 
\begin{equation}
  0 = C (1-\Lambda_z^2)^2 + 2 \Lambda_z  (1+\Lambda_z^2)
  \; .
\end{equation}
A similar equation can be obtained for $\VECLambda=\Lambda_x\VECx$. We observe that the total force acting on the membrane due to the container is only equilibrated for special values of $\VECLambda$. This implies that the conformal transformation defined in Eq.~(\ref{eq:conformaltransformationconfinedgeneral}) does not yield an equilibrated surface in general. However, we have shown in Sec.~\ref{sec:confinednonaxisymmetricsolutions} that these surfaces can be used as initial conditions for subsequent finite element simulations in order to obtain equilibrium shapes.

%%%%%%%%%%%%%%%%%%%%%%%%%%%%%%%%%%%%%%%%%%%%%%%%%%%%%%%%%%%%%%%%%%%%%%%%%%%%%%%%

\section{Finite element simulations \label{app:finiteelementsimulations}}
In the simulations the fluid membrane is discretized by a set of triangles with $N\sim 4000$ nodes using subdivision finite elements \cite{Klug2006,Klug2008}. The position of the surface in $\RR^3$ is parametrized with the local coordinates $s_1$ and $s_2$ and interpolated by 
\begin{equation}
  \VECx_h (s^1,s^2) = \sum_{a=1}^N \VECx_a N^a (s^1,s^2)
  \; ,
\end{equation}
where $\VECx_a$ is the position of node $a$ and $N^a$ are the Loop subdivision trial functions \cite{Cirak}.
We search for equilibrium solutions of the energy functional
\begin{equation}
  E = \int  \left[ 2 H^2 + \frac{\mu_A}{2} (\sqrt{a} - \sqrt{\bar{a}})^2  \right] \, \romd s^1 \romd s^2  +  \frac{\mu_V}{2} (V - \bar{V})^2
  \; ,
\end{equation}
where the constants are set to $\mu_A = 10^5$ and $\mu_V = 5 \cdot 10^4$ to ensure that the membrane adopts the prescribed target values for area and volume to a numerical error of about $10^{-3}$.  Note that the area constraint is recast in a local form (\textit{local area constraint method}), in which $\sqrt{a}$ is the surface Jacobian such that $A= \int \sqrt{a} \, \romd s^1 \romd s^2$ \cite{Klug2006,Klug2008}.

The energy can be expressed in terms of the nodal positions $\VECx_a$ to determine the gradient with respect to $\VECx^a$. This yields the corresponding nodal forces $\VECf_a^{M}$ due to the elasticity of the membrane (for more details see appendix~B of Ref.~\cite{OsmanNorbertMartin2012B}). 
The contact between the membrane and the spherical container is modeled by a harmonic force which is only applied to the nodes which leave the container. The force acting on such a node $a$ is given by
\begin{equation}
  \VECf_a^C = k d_a^2 \VECn
  \; ,
\end{equation}
where $k= 1.5 \cdot 10^6$ is the stiffness constant, $d_a$ the penetration depth, and $\VECn$ the inward-pointing normal of the container. 
The total force at node $a$ is the sum of all contributions 
\begin{equation}
  \VECf_a = \VECf_a^{M} + \VECf_a^{C}
  \; .
\end{equation}
To find an equilibrium solution, we integrate these nodal forces $\VECf_a$ in time according to Newton's equations of motion until a balance of forces is reached, \textit{i.e.}, until $\VECf_a=0$  (for more details we refer again to appendix~B of Ref.~\cite{OsmanNorbertMartin2012B}).

All our input meshes are obtained from the axisymmetric Runge-Kutta solutions (see appendix~\ref{app:RungeKutta}). Every cross section of such a solution, which contains the symmetry axis, yields the same profile regardless of the azimuthal angle $\phi$. In order to obtain a three-dimensional mesh we discretize this profile in a way which does not depend on $\phi$. We first choose the angular stepsize $\Delta\phi$, \textit{i.e.}, the angle between two adjacent planes. Then we establish a meridional stepsize, \textit{i.e.}, the distance between two successive points in the same plane. For the first point, we calculate the distance between this point and the corresponding first point in the adjacent plane $d_{1\to 2}=\rho_1\Delta\phi$. This distance is taken as the stepsize between the first and the second point of the same plane. We repeat this operation for all the points $i$, $d_{i \to i+1}=\rho_i\Delta\phi$ until we reach the first point again to close the profile.
We obtain a mesh composed of squares with edges that are the bonds between two successive points of the same plane and the bonds between two analogous points (of same arc length) in two adjacent planes. Dividing these squares into two we obtain triangles that are almost equilateral. Moreover, they have the advantageous property that they are smaller close to the symmetry axis where the curvature becomes larger. This property is important in order to increase the precision without increasing the number of the nodes dramatically. 
For the non-axisymmetric meshes we apply a conformal transformation to each node of the axisymmetric mesh while keeping the same connectivities. In this way we preserve the proportionality between the curvature and the size of the triangles.

%%%%%%%%%%%%%%%%%%%%%%%%%%%%%%%%%%%%%%%%%%%%%%%%%%%%%%%%%%%%%%%%%%%%%%%%%%%%%%%%

\end{document}